\documentclass[aps,prd,reprint,superscriptaddress,floatfix,twocolumn,nofootinbib]{revtex4-2}

\usepackage{xcolor,amssymb,amsmath,multirow,epsfig,graphicx,color,booktabs,cancel}

\newcommand{\delete}[1]{\textcolor{red}{\st{#1}}}

\usepackage[normalem]{ulem}
\usepackage{soul}
\usepackage{mathtools}
\usepackage{mathrsfs}
\usepackage{bm}
\usepackage{relsize}
\usepackage{indentfirst}
\def\be{\begin{eqnarray} &&}
\def\nonu{\nonumber \\ &&}

\def\ee{\end{eqnarray}}

\def\bwt{\begin{widetext}}
\def\ewt{\end{widetext}}

\newcommand{\bpmat}{\begin{pmatrix}}
\newcommand{\epmat}{\end{pmatrix}}
\setbox0=\hbox{$%
\xdef\fdfourteen{\the\fontdimen14\textfont2}
\xdef\fdfifteen{\the\fontdimen15\textfont2}
\xdef\fdsixteen{\the\fontdimen16\textfont2}
\xdef\fdseventeen{\the\fontdimen17\textfont2}
$}
\newcommand\newss{\setbox0=\hbox{\(\)}%
\fontdimen14\textfont2=1.8ex
\fontdimen15\textfont2=1.8ex
\fontdimen16\textfont2=1.1ex
\fontdimen17\textfont2=1.1ex
}

\begin{document}

\title{Solving the homogeneous Bethe-Salpeter equation with a quantum annealer}

\author{Filippo Fornetti}
\affiliation{Dipartimento di Fisica e Geologia, Universit\`a di Perugia, and
 INFN, Sezione di Perugia, Via Alessandro Pascoli,
06123 Perugia, Italy}
\author{Alex Gnech}
\affiliation{Department of Physics, Old Dominion University, Norfolk, Virginia 23529, USA and
Theory Center, Jefferson Laboratory, Newport News, Virginia 23606, USA}
\author{Tobias Frederico}
\affiliation{Instituto Tecnol\'ogico de Aeron\'autica, 12228-900 S\~ao Jos\'e dos Campos,
Brazil}

 \author{Francesco Pederiva}
\affiliation{Dipartimento di Fisica,  Universit\`a di Trento, and INFN-TIFPA, Via Sommarive 14, 
Povo 38123, Trento, Italy}

\author{Matteo Rinaldi}
\affiliation{INFN, Sezione di Perugia, Via Alessandro Pascoli,
06123 Perugia, Italy and
Dipartimento di Fisica e Geologia, Universit\`a di Perugia, Italy
 }
\author{Alessandro Roggero}
\affiliation{Dipartimento di Fisica,  Universit\`a di Trento, and INFN-TIFPA, Via Sommarive 14, 
Povo 38123, Trento, Italy}

\author{Giovanni Salm\`e}
\affiliation{
INFN, Sezione di Roma, P.le A. Moro 2, 00185 Rome, Italy}

\author{Sergio Scopetta}
\email[S. Scopetta is presently Science Counsellor at the Italian Embassy in Spain, calle Lagasca 98, 28006 Madrid, Spain.]{  sergio.scopetta@esteri.it}
\affiliation{Dipartimento di Fisica e Geologia, Universit\`a di Perugia, and
 INFN, Sezione di Perugia, Via Alessandro Pascoli,
06123 Perugia, Italy}

\author{Michele Viviani}
\affiliation{INFN, Sezione di Pisa,
Largo Pontecorvo 3, 56100, Pisa, Italy}

\date{\today}

\begin{abstract}
The homogeneous Bethe-Salpeter equation (hBSE),  describing a bound system in a genuinely relativistic quantum-field theory framework, was solved for the first time by using a  D-Wave quantum annealer.  After applying standard techniques of discretization, the hBSE, in ladder approximation, can be formally transformed in a generalized eigenvalue problem (GEVP), with two square matrices: one symmetric and the other non symmetric. The latter matrix poses the challenge of obtaining   a suitable formal approach for investigating the {\em non symmetric} GEVP    by means of a quantum annealer, i.e to recast it as a quadratic unconstrained binary optimization  problem. A broad numerical analysis of the proposed algorithms, applied   to  matrices of dimension up to 64,  was carried out by using both the proprietary simulated-annealing  package  and the D-Wave {\em Advantage 4.1 system}. The numerical results very nicely  compare with  those obtained with standard classical algorithms, and also show interesting  scalability features. 
\end{abstract}
\maketitle
\section{Introduction}
\label{sect_intro}
The past decade has seen  an exponential growth of  research activity and interest in quantum computing in an impressive  number of  fields  (see, e.g.,  Refs. \cite{Preskill:2018jim,Preskill:2022kgz,Bauer:2022hpo,Funcke_2023,delgado2022quantum,cao2019quantum,bauer2020quantum,emani2021quantum,flother2023state,bova:2021,herman2022survey} for recent reviews which could offer an admittedly partial idea of the state-of-the-art). The  driving force behind all of this  has been the  astonishing advances in  new quantum-hardware implementation  which occurs with close frequency, as shown by the long list of major achievements  one can find in literature and  media (see, e.g. Ref. \cite{Cheng:2023csj} for a an introduction of the  main  quantum-computing platforms, as well as algorithms).
 However, one should recall that an easy path toward a universal, fault-tolerant quantum computer is currently hindered by noise, decoherence   and scalability \cite{Preskill:2018jim}.

 
In general, one can perform   calculations that take advantage of  either digital gate-based quantum computing   or analog quantum annealing (see, e.g., Ref. \cite{Alexeev:2020xrq} for a detailed discussion of the two approaches).
The first choice is more widely adopted, since it is meant as the quantum analog of a classical  general-purpose computer, and  it has the possibility to address a  broader set of problems (see, e.g., Ref. \cite{kim2023evidence}, for  recent experimental results obtained  using an IBM quantum machine).
 The second possibility, the quantum annealing,   is mainly devoted to optimization problems, that can be properly encoded into an Ising Hamiltonian. In this case, one  searches for the ground-state of the system by exploiting  quantum fluctuations  (see, e.g., Refs.~\cite{finnila1994quantum,kadowaki1998quantum,morita2008mathematical,santoro2006optimization,Das:2008zzd}  for an introduction to the foundations of  quantum annealing).   Quantum annealing,  that has its basis  in the adiabatic theorem,  have been exploited  in several practical applications (see, e.g., Refs.\cite{baldassi2018efficiency,hauke2020perspectives,Abel:2021fpn,King:2022okf,rajak2023quantum} and references quoted therein for a non exhaustive list of quantum-annealing applications) outperforming the results obtained  with simple classical heuristics, like simulated annealing. 
In this work we exploit this second possibility to solve the {\it non symmetric} Generalized EigenValue Problem (GEVP). In particular we use the D-wave Advantage quantum annealer (\texttt{QA}) \cite{mcgeoch2020d,mcgeoch2021advantage,mcgeoch2022advantage} which has a network of  more than 5000 qubits, with 15-way connectivity.


 The application of the quantum annealer for finding the solution of the GEVP for {\em symmetric} matrices it is very well known in literature~ \cite{Krak:2021,krakoff2022controlled,Illa:2022jqb,tosti2022review,stockley2023optimizing}.
In this work,  we aim instead to find the minimal/maximum right eigenvalue (and corresponding eigenvector) of a {\em non-symmetric} GEVP by the use of a quantum annealer. Notice that the proposed technique  could also be used for the left eigenvectors with simple modifications.
In our specific case we focus on the {\em non symmetric} GEVP stemming from the discretization method adopted for solving   the homogeneous  Bethe-Salpeter equation  (hBSE) \cite{BS51,GellMann:1951rw}  in ladder approximation,  directly in the four-dimensional (4D) Minkowski space (see also Refs. \cite{itzykson2012quantum,Nakanishi:1971} for a detailed introduction to the topic, and the brief resum\'e in Appendix \ref{app_hBSE}). 
In particular we focus on the numerical solutions of the hBSE for two massive scalar fields interacting through the ladder exchange of another massive scalar field (see Ref. \cite{FSV2} for the results obtained with a classical computation). 

The {\it non symmetric} GEVP is  given in general by the expression
\be \label{eq:ns_gevp}
A~{\bf v}_i=\lambda_i~B~{\bf v}_i
\label{eq:gevp}\ee
where  the $n\times n$ matrix  $A$ is a real and  {\em non symmetric}, the $n\times n$ matrix   $B$  is non singular and {\em symmetric} and, ${\bf v}_i$  is the eigenvector corresponding to the i-th eigenvalue $\lambda_i$. 
It is important to underline that this form  is relevant for many areas, e.g., for the analysis of mechanical systems \cite{alliney1992variational,KIM2007950},  in fluid mechanics \cite{li2004eigenvalue},  for  investigating  criticality in nuclear reactors \cite{scheben2011iterative} or the onset of oscillatory instability in a heated cavity \cite{burroughs2004linear}, for lattice QCD \cite{Orginos:2015tha}, in industry \cite{Yarkoni:2021zvu}, etc.  

At the present stage,  we reduce the GEVP into a standard eigenvalue problem  by inverting the matrix $B$ through a classical algorithm, before passing it to the quantum annealer to  compute the real maximal/minimal eigenvalue and eigenvector. 
 This is a crucial first step of the investigation since it permits to asses the
 viability of this algorithm before being generalized.

The problem is then divided in two steps.
The first step is to translate the GEVP into a quadratic unconstrained binary optimization\footnote{One aims to get a quadratic polynomial over binary variables.} (QUBO), constructing the suitable {\em objective function} (OF)  in terms of binary variables. The second step, carried out through the  proprietary software, is to  map the QUBO problem onto such  Ising-model, embodied by the qubit architecture (see Ref.~\cite{johnson2011quantum} for the first implementation of a QUBO problem on  a D-wave \texttt{QA}  and Refs. \cite{mcgeoch2020d,mcgeoch2021advantage,mcgeoch2022advantage} for the D-Wave Advantage system). The quantum annealear is then used  to search for the global minimum of the corresponding  Ising Hamiltonian implemented on the physical quantum processing units (QPUs).


 The paper is organized as follows. In Sec. \ref{sect_general}, the physics problem  from where the {\em non symmetric} GEVP stemmed from is briefly illustrated.  In Sec. \ref{sect_objf}, the OF, which our GEVP is mapped onto, and its expression in terms of the binary variables are presented in detail. In Sec. \ref{sect_results}, the results of our numerical investigation on the annealing simulator, \texttt{SA}, as well as the D-wave Advantage System \texttt{QA} are discussed at some length and compared. Finally, in Sec. \ref{sect_conclusions},  our conclusions are drawn with a focus on  the  perspectives. Appendices are devoted to supply more details on the physics case and the construction of the OF.
\section{The physics problem}
\label{sect_general}
As previously stated, our aim is to  solve    the homogeneous Bethe-Salpeter equation ~\cite{BS51} of a system composed by two massive scalars bound through the ladder exchange of a third massive scalar, 
directly in Minkowski space. A  variational method, tailored for a \texttt{QA}, is applied in order to obtain the ground-state of the system. This     non-perturbative  quantum-field-theory problem was solved in Ref. \cite{FSV2} by using the Nakanishi integral representation \cite{nak63} (NIR)  of the Bethe-Salpeter (BS)  amplitude, i.e. the key quantity for describing the ground-state of the  system and eventually calculating relevant observables. Following Ref. \cite{FSV2}, one can formally transform the hBSE in Minkowski space into a GEVP  adopting a standard discretization method. In particular,   a  polynomial ortho-normal basis, composed by the Cartesian product of Laguerre and Gegenbauer polynomials, was adopted. It should be pointed out that   solutions obtained directly in Minkowski space make a more direct comparison between the theoretical outcomes and the experimental results possible, without resorting to Euclidean solutions  (see, e.g.,  Ref. \cite{Roberts:1994dr,Maris:2003vk,Bashir:2012fs} for an introduction to the Dyson-Schwinger plus hBSE framework, in Euclidean space). Let us recall that  Euclidean solutions  could be directly adopted for  describing physical observables, once  the assumptions of the famous theorems by Osterwalder and Schrader \cite{Osterwalder:1973dx,Osterwalder:1974tc}, stating  necessary and sufficient conditions
for formally bridging   Euclidean and Minkowskian quantum-field theory,  are fulfilled and the set of unavoidable approximations is actually under control.

From a bird's eye view,   the two-body BS amplitude  allows to reconstruct  the residue of the four-leg Green's function at the bound-state mass pole. Moreover,   one can show that such an amplitude is the solution of  a homogeneous integral equation, i.e. the hBSE.  Notice that, in relativistic quantum-field theory, the hBSE has the same role in the bound-state description that  the Schr\"odinger equation  has in non relativistic quantum-mechanics.

In conclusion, one is able 
to  formally transform the Minkowskian hBSE into the {\it non symmetric}
GEVP  of the form in Eq.\eqref{eq:ns_gevp}. {The eigenvalue is proportional to the inverse of the square coupling constant present in the interaction vertex (produced by the  underlying field theory). Since the coupling constant is real  for physical reasons,    real eigenvalues must be sought.} It should  be pointed out that the eigenvalues and corresponding eigenvectors  (eigenpair) in a {\em non symmetric} GEVP can be real or complex. In case they are complex  they appear in conjugated pairs. Moreover, the eigenvectors ${\bf v}_i$   do not have the property to be orthogonal.
{As specified above, in} our case we are interested only in the real  eigenvalues, and particularly the one  corresponding to the largest, positive real $\lambda_i$, since  it is the inverse of the minimal coupling constant that allows the existence of a bound system with a given mass \cite{FSV2}.

\section{\texttt{QA} modeling of a non-symmetric generalized eigenvalue  problem}
\label{sect_objf}
 In order to perform numerical calculations on  a \texttt{QA},   the actual problem has to be expressed as a
QUBO model  or, equivalently, to find 
an  Ising-model form  for the problem under scrutiny
(see, e.g., Ref. \cite{glover2018tutorial} for a tutorial on formulating and using QUBO models). 

For a symmetric GEVP, where the eigenvalues are real, 
 there are already   several
investigations with specific application to different problems (see, e.g., Refs.  \cite{Teplu:2019,pastorello2019quantum,Teplu:2020,
krakoff2022controlled,Krak:2021,Illa:2022jqb}). In all  these applications,
the Rayleigh-Ritz variational principle is adopted to transform the eigenvalue problem in a QUBO form.
For the {\em non symmetric} GEVP, involving also complex
eigenpairs,    dedicated efforts are  still needed.
Recalling that QUBO problems are by definition strictly symmetric,  one should be able to replace  the non symmetric original problem with a symmetric one, which allows to evaluate the set of real eigenpairs we are interested in.
This will be achieved by transforming the initial   GEVP into a symmetric one with an equivalent spectrum.  Thus,  the same algorithm used for applying the Rayleigh-Ritz variational principle, with minimal changes, can be adopted.  In performing such a transformation, the minimization problem acquires a quadratic dependence on the parameter playing the role of eigenvalue. This means that each eigenvalue is a minimum.  This on one side can be usefully exploited for addressing  the entire  spectrum on the other side require to guide the search toward the relevant region where the eigenvalue has to  be found. 

In this Section we go through the formalism we adopted for the non-symmetric case with some detail.

\subsection{Formalism}
Since the  physical case under consideration generates a non-symmetric matrix $A$ in Eq. \eqref{eq:ns_gevp},
 the   Rayleigh-Ritz variational principle is not directly exploitable. 
 A first trivial attempt to overcome this problem might  be to single out  the symmetric part of the matrix $A$, and  solve the corresponding symmetric GEVP, recalling that $B$ by itself is symmetric. Unfortunately, this formal manipulation  does not  return the original global minimum. This is illustrated in Appendix \ref{app_nseig},
 where an inequality between the eigenvalues of a non-symmetric matrix and the ones of its symmetric part is discussed in the simpler case of a standard eigenvalue problem. Therefore,   the eigenvalue problem $\Bigl[(A+A^T)/2\Bigr]  ~{\bf v}_i= \lambda_i~B ~{\bf v}_i$ cannot be used for determining the minimal/maximal eigenvalue of Eq. \eqref{eq:ns_gevp}. The same issue is met also for other  methods of symmetrization, like $A^TA$    
 (see  Appendix \ref{app_nseig} for more discussion and examples). 

A possible way to overcome this problem was discussed in Ref.~\cite{alliney1992variational}, where 
the symmetrization of the original problem is recast as  the product of $ A-\tilde \lambda~B$ with its transpose, i.e.
\be
{ S}=\Bigl[A^T-\tilde\lambda~B^T\Bigr]
  \Bigl[A-\tilde\lambda~B\Bigr]\,,
\label{eq:sym}
\ee
where $\tilde \lambda$ can be  complex (see below).

The resulting OF, which now contains a symmetric product, reads
\be
 f(A,B,{\bf v},\tilde\lambda)={\bf v}^T ~\Bigl[A^T-\tilde\lambda~B^T\Bigr]
  \Bigl[A-\tilde\lambda~B\Bigr]~{\bf v}=
 \nonu
 ={\bf v}^T ~\Bigl[A^TA -2\tilde \lambda M +\tilde\lambda^2 B^T B\Bigr]~{\bf v}\ge 0 ~~,
 \label{eq:objf}\ee
 where  ${\bf v}$ is chosen to be a {\em real} normalized vector  and   $M$ is a symmetric matrix, defined as:
 \be 
 M={(B^T A+A^T B)\over 2}~~.
 \label{eq:defM}
 \ee
In what follows, it should be kept in mind that our goal is to find a real eigenpair of the analyzed GEVP, although the OF $f(A,B,{\bf v},\tilde\lambda)$ vanishes also for complex eigenpairs.
 
 The striking difference between   the global-minimum search of  $f(A,B,{\bf v},\tilde\lambda)$ and  the standard eigenvalue problem is the quadratic dependence upon the variable $\tilde \lambda$.
 Calling  $\lambda$ the solution of  the quadratic form in Eq.~\eqref{eq:objf} for a given  real vector ${\bf v}$, one gets  
 \be
\lambda({\bf v})= \lambda^R({\bf v})\pm i\lambda^I({\bf v})
\ee
where the real part is
\be
\lambda^R({\bf v})= {\bar M({\bf v})\over  |B~{\bf v}|^2}~.
\label{eq:lambda_r}
\ee 
with $
 \bar M({\bf v})={\bf v}^T ~M ~{\bf v}$. 
  The imaginary part is written as follows
 \be
 \lambda^I({\bf v})={\sqrt{{\cal I}({\bf v})}~  \over  |B~{\bf v}|^2}
 \ee
 where 
 \be
 {\cal I}({\bf v})=||A~{\bf v}||^2~ ||B~{\bf v}||^2- \bar M^2({\bf v})~~,
 \label{eq:Rv}\ee
 with $ ||A~{\bf v}||^2={\bf v}^T ~A^TA~{\bf v}$ and   equivalently for $ ||B~{\bf v}||^2$. Notice  that ${\cal I}({\bf v})$ is a positive quantity, since
\be 
 {\cal I}({\bf v})=||A~{\bf v}||^2~ ||B~{\bf v}||^2-\left[{\bf v}^T ~{B^T A+A^TB\over 2}{\bf v}
 \right]^2 
 \nonu\ge ||A~{\bf v}||^2~ ||B~{\bf v}||^2-\left[{||{\bf v}^T B^T A {\bf v}||
 +||{\bf v}^TA^TB{\bf v}||\over 2}
 \right]^2
 \nonu \ge   ||A~{\bf v}||^2~ ||B~{\bf v}||^2-||{\bf v}^T B^T A {\bf v}||^2
 \ge 0~.
 \label{eq:Rx_gep}\ee
Therefore,  our task is to set up an algorithm for finding a real vector ${\bf v}$ that minimizes both the 
OF and   the imaginary part  of the eigenvalue, $\lambda^I({\bf v})$, so that the set $\{{\bf v},\lambda^R({\bf v})\}$ is the searched eigenpair.
Notice that a real vector ${\bf v}$ that minimizes ${\cal I}({\bf v})$, also minimizes the OF $f(A,B,{\bf v},\tilde\lambda= \lambda^R)$. However, the opposite is not true, since   the OF vanishes also for complex eigenpairs. 
 
In Ref. \cite{alliney1992variational}, the search was carried out by classically minimizing ${\cal I}({\bf v})$ through a gradient-descent method. Unfortunately, ${\cal I}({\bf v})$ has a quartic dependence on ${\bf v}$, and therefore the corresponding minimization cannot straightforwardly be translated into a QUBO problem. The quartic problem can  be recast into a quadratic one by  adding suitable penalties,  at the price of a substantial increase of  the dimension of the QUBO problem   (see, e.g., Ref. \cite{Chang:2018uoc,glover2018tutorial}).  However this will require to substantially increase the matrix size to pass to the annealer. In order to overcome the difficulties, we approached the problem in a different way. 

First of all,  given the exploratory nature of our investigation, instead of directly using the OF in Eq.~\eqref{eq:objf}, we introduced a hybrid algorithm, namely, we  transformed the initial GEVP  into a standard one, leaving the generalized problem for future investigations.
We exploit the   symmetry and non singularity of the matrix  $B$, decomposing it by using the  standard $LDL^T$ factorization, with $L$  a lower triangular matrix, having diagonal elements equal to $1$, and $D$  a non singular diagonal matrix (see, e.g., the \texttt{LAPACK} library). Hence,  the  original GEVP, Eq. \eqref{eq:ns_gevp}, becomes
\be
C~L^T{\bf v}_i=  \lambda_i ~L^T{\bf v}_i~,
\label{eq:LDLT}\ee
where \be 
C=D^{-1}L^{-1}~A~[L^T]^{-1}~~,
\label{eq:C_def}\ee
is non symmetric, and the real vector $L^T{\bf v}_i$ is no longer normalized. Since the global-minimum search is not affected by multiplicative factors, we can use a normalized vector formally defined by    
\be 
{\bf w}=L^T{\bf v}/||L^T{\bf v}|| ~.
\label{eq:w_v}
\ee
Then, one can address the QUBO problem by replacing
  the initial  OF
 with  an equivalent one, given by  
\be\label{eq:objfc}
f(C,{\bf w},\tilde\lambda)= {\bf w}^{ T}  ~\Bigl[C^T-\tilde\lambda\Bigr]
  \Bigl[C-\tilde\lambda\Bigr]~{\bf w}~.
\label{eq:newobj}\ee 
The real part of the solution of the quadratic form in $\tilde \lambda$, for a given ${\bf w}$, is 
  \be
  \lambda^R({\bf w})={1\over |{\bf w}|^2}~{\bf w}^{ T}   {C^T+C\over 2} {\bf w}
   ~,
   \label{eq:lambda_rn}
  \ee
  while the imaginary part reads
  \be
  \lambda^I({\bf w})=\sqrt{{|C {\bf w}|^2\over |{\bf w}|^2}-\Bigl[\lambda^R({\bf w})\Bigr]^2}
   ~~.
  \label{eq:lambda_in}
  \ee
  Since we have chosen ${\bf w}^T{\bf w}=1$,  it is understood that in order to  get the normalized eigenvector of the original GEVP, one has to  apply the inverse transformation to ${\bf w}$ and normalize the result, i.e.
  \be
  {\bf v}= [L^T]^{-1}~{\bf w}/||[L^T]^{-1}~{\bf w}||~.
  \ee

\subsection{Algorithm implementation}
The OF in Eq. \eqref{eq:newobj} is explicitly written as follows:
\be\label{eq:objfc1}
f(C,{\bf w},\tilde\lambda)={\bf w}^T {\cal S}(\tilde \lambda) {\bf w}=\sum_{i,j}w_i w_j {\cal S}_{i,j}\,,
\ee
where ${\cal S}$ is  an  $n\times n$ matrix  ($n$ is dictated by the discretization method applied to the hBSE) given by 
\be
{\cal S}_{i,j}(\tilde \lambda)=\tilde \lambda^2\delta_{i,j}-\tilde \lambda (C_{i,j}+C_{j,i})+\sum_\ell C_{\ell,i}C_{\ell,j}\,,
\label{eq:sym_mat}\ee
and ${\bf w}\equiv\{w_i\}$ is a normalized vector. By using the binary representation introduced in  Appendix \ref{app_bb},  the components $w_i\in[-1,1]$  are expressed in terms of the  binary basis as follows 
\be
w_i=-q_{i,b}+\sum_{\ell=1}^{b-1} \frac{q_{i,\ell}}{2^\ell}~,
\label{eq:bin_exp}\ee
where $b$ is the number of bits, $q_{i,\ell}=0,1$ and  the bit $q_{i,b}$  carries the sign, i.e.   for 
$w_i\ge 0$ ($w_i<0$) one assigns $q_{i,b}=0$ ($q_{i,b}=1$). Notice that the binary expression in Eq. \eqref{eq:bin_exp} belongs to the interval $[-1,1)$.

The string $\{-1,1/2,1/4,~\dots~, 1/2^{b-1}\}$ is  the  so-called  {\em precision vector}, ${\bf p}$,  since the last term $1/2^{b-1}$ controls the precision at which   $w_i\in[-1,1)$ is approximated. Within our notation ${\bf p}$ is a column vector with dimension $b$. Using the precision vector ${\bf p}$, it is possible to construct a rectangular matrix, $P^T$, with $n$ rows and $n\times b$ columns, that allows to transform  a  real vector ${\bf w}$ into its binary expression,  and eventually 
$f(C,{\bf w},\tilde\lambda)$  in a form suitable for a QUBO evaluation by using a \texttt{QA}. In a compact form, one writes
\be
{\bf w}= P^T~ {\bf x}~,
\label{eq:vect_bin}\ee
where
\be
P^T=\bpmat ~{\bf p}^T  &     0    &\dots   & 0
         \\ 0      &~{\bf p}^T& \dots  & 0
         \\ \vdots &\vdots    & \ddots & 0
         \\ 0 &0    & \dots  & {\bf p}^T\epmat~,
\ee
and
\be
{\bf x}=\bpmat {\bf q}_{1;b}  \\ 
          {\bf q}_{2;b}  \\ 
          \vdots          \\ 
          {\bf q}_{n;b} \epmat~,
\ee
with ${\bf q}_{i;b}\equiv \{q_{i,1},q_{i,2},\dots,q_{i,b}\}$.
Inserting Eq.~(\ref{eq:vect_bin}) in  Eq.~(\ref{eq:objfc1}) one obtains (see Appendix \ref{app_bb} for details)
\be
f(C,{\bf w},\tilde\lambda)={\bf x}^T ~P{\cal S}(\tilde \lambda) P^T~{\bf x}
~.
\label{eq:objfc_bin}\ee
One easily recognizes that ${\bf x}$ is a column vector containing $n\times b$ bits, leading to the expected QUBO form of our problem, and  $P{\cal S}(\tilde \lambda)P^T$ is a   square symmetric matrix, with    dimensions $(n\times b)~\times~(n\times b)$. 

Starting with some given real value of $\tilde \lambda$ (see the next subsection), that  guides 
the algorithm towards a real-valued eigenpair,  the \texttt{SA} or the \texttt{QA}  returns  a binary vector ${\bf x}$ for each annealing cycle $N_A$.
Among the $N_A$ vectors ${\bf x}$ we select the one that gives the minimal value of $f(C,{\bf w}, \tilde \lambda)$. Such a vector allows to get  ${\bf w}$, necessary for     calculating $\lambda^R({\bf w})$   and  $\lambda^I({\bf w})$, using  Eqs. \eqref{eq:lambda_rn} and   \eqref{eq:lambda_in}, respectively.
To accurately determine ${\bf w}$, which minimizes at the same time the number of bits required in the \texttt{QA}, a two-step search is implemented: 
the first step is the  guess phase (GP), and the second one is the iterative gradient-descent phase (DP) (see, e.g.,  Refs. \cite{Krak:2021,krakoff2022controlled} and \cite{Illa:2022jqb}  for the symmetric case).

\subsection{Initial guess for the non-symmetric case}
\label{sect_ig}
Differently from what happens for the symmetric case \cite{Krak:2021,krakoff2022controlled,Illa:2022jqb,stockley2023optimizing}, the non-symmetric one leads to a non linear dependence on the scalar parameter $\tilde \lambda$, and therefore the  search needs a particular care. It should be emphasized that 
the OF is minimized by the {\em entire} set of real eigenpairs. Hence, in order  to point to the largest, positive  real eigenvalue, we need to add further information  to our algorithm. We can  guide the search of the minimum by using the Gershgorin circle theorem (see, e.g.,  Ref.~\cite{wolkowicz1980bounds,varga2010gervsgorin}, for an introduction). Noteworthy, it  provides  bounds for   each eigenvalue of a given complex matrix, allowing the search range to be narrowed.

The theorem yields
\be
|\lambda -a_{ii}|= \sum_{j\ne i} \left| {a_{ij} x_j(\lambda)\over x_i(\lambda)}\right|\le  \sum_{j\ne i} \left| a_{ij}\right|=R_G(a_{ii})~,
\ee
where $\lambda$ is an  eigenvalue of the given matrix, with elements $a_{ij}$, { and  $x_{i(j)}(\lambda)$ } are  components of the corresponding eigenvector, 
{ such that $|x_j/x_i|\le 1$, $\forall ~j \ne i$}. Hence, an eigenvalue belongs to a proper disc, with radius $R_G(a_{ii})$ and  center $a_{ii}$. The same is valid by summing over the columns, once  the transpose is considered. 
It is also worth mentioning a second Gershgorin theorem \cite{varga2010gervsgorin}. It  proves  that  if a disc is  disconnected from all  the others, then it contains  one and only one eigenvalue, which necessarily  is real (the complex ones are conjugated). 

Inspired by the Gershgorin theorem, a   suitable permutation can be applied in order to rearrange the matrix $ C$ in Eq. \eqref{eq:C_def}, so that  the diagonal elements are in  decreasing order, i.e. $ c_{11}\ge c_{22}\ge \dots \ge  c_{(n\times b)(n\times b)}$. 
%
%
The  initial value  of $\tilde \lambda$ is chosen  equal to $c_{11}$. The matrix $P{\cal S}(\tilde \lambda)P^T$ is passed to the \texttt{QA} or the  \texttt{SA}. For each of the $N^{GP}_A$  annealing cycles the  binary vectors ${\bf x}_\alpha$, and the corresponding values of the OF is returned. With $\alpha$ we label the annealing cycle.
Following the Gershgorin constraint, we  analyze the set
$$\{ {\bf x}_\alpha, f(C,{\bf w}_\alpha,\tilde\lambda=c_{11}), \lambda^R({\bf w}_\alpha),\lambda^I({\bf w}_\alpha)\}~,$$
and   eliminate the solutions such that  $$|\lambda^R({\bf w}_\alpha)-c_{11}|>R_G(c_{11})~.$$
Among the surviving solutions,   the one that satisfies the condition
\be 
{f}_{best}^{GP}=\min_{{\bf w}_\alpha} f(C,{\bf w}_\alpha,\tilde\lambda=\lambda^R({\bf w}_\alpha))
\label{eq:min_GP}\ee
is retained, and $\lambda^{ R}(\bf{w}_{\alpha})$ is calculated. The minimization search is repeated three times, $i=1,2,3$,   testing if the condition ${f}^{GP}_{best;i} < {f}^{GP}_{best;i-1}$ is satisfied.  If there is an improvement, the next run receives $\tilde \lambda= \lambda^{ R}(\bf{w}_{\alpha_{best;i}})$, otherwise the iteration stops and the GP is closed.  Empirically, we found this procedure is more effective in finding a good initial eigenvector than just increasing the number of annealing samples in a single iteration.
After completing the minimization search, the best pair, i.e. $\lambda^{\bf R}(\bf{w}_{{GP}})$ and the corresponding real vector ${\bf w}_{{GP}}$, that satisfies the condition in Eq.~(\ref{eq:min_GP}), is passed to the gradient-descent phase. The algorithm of the GP is schematically illustrated in Tab. \ref{tab:guess_phase}, and 
 the numerical results that illustrate the behavior of the algorithm are discussed in  Section~\ref{sec:simulation:gp}.
\begin{table*}[htb]
\renewcommand{\arraystretch}{1.5}
\caption{ The  guess-phase search  of  the minimum for the OF in Eq. \eqref{eq:objfc1}, obtained from Eq. \eqref{eq:objf} after applying a classical $LDL^T$ factorization. An assigned number of bits $b$  is used for the transformation to the QUBO form, Eq. \eqref{eq:objfc_bin} (see text).  }
    \centering
    \begin{tabular}{c l}
   \toprule 
   ~& {\bf Algorithm  for the guess-phase, with a given number of bits $b$}\\
    \midrule 
        1:~ & {\bf Assign} the input value  $\tilde \lambda=c_{11}$,   suggested by  the Gershgorin theorem \\
        2: ~& {\bf While} ${f}^{GP}_{best;i} < {f}^{GP}_{best;i-1}$, for $i=1,2,3$\\
        3:~ &  \hspace{0.5cm} Minimize the QUBO form, Eq. \eqref{eq:objfc_bin}, with  $N_A^{GP}$ annealing cycles\\
       4: ~&  \hspace{0.5cm}{ Look  for} acceptable  $\lambda^R({\bf w}_{\alpha_i})$ falling in the Gershgorin disc\\
     5: ~& \hspace{0.5cm} Among the acceptable solutions,  check
     if ${f}^{GP}_{best;i} < {f}^{GP}_{best;i-1}$ and    continue the loop,  replacing $\tilde{\lambda}$ by $\lambda^{ R}({\bf w}_{\alpha_{best;i}})$  \\
     6:~ &  Put $  \lambda^{ R}({\bf w}_{{GP}})=\lambda^{ R}({\bf w}_{\alpha_{best;i}})$ and ${\bf w}_{{GP}}={\bf w}_{\alpha_{best;i}}$ and pass those quantity to  the gradient-descent phase\\
 \bottomrule
    \end{tabular}
  \label{tab:guess_phase}
\end{table*}
\subsection{Gradient-descent phase}
\label{sec:gradient_dm}
The algorithm proceeds through a second phase: the gradient-descent phase, which is   based on the Hessian of the OF in Eq. \eqref{eq:objfc1}. Let us call ${\bm \delta}(z)\equiv\{\delta_i(z)\}$ the variation of the vector ${\bf w}^z$, i.e. ${\bm \delta}(z)={\bf w}^z-{\bf w}^{z-1}$, where $z$ labels the  iterations or zoom steps of the gradient-descent phase.   Note that ${\bf w}^0={\bf w}_{{GP}}$. Then, the  following binary expression for the ${\bm \delta}(z)$ components  holds
\be
\delta_i(z)= {1\over 2^z}~ \Biggl(-q_{i,b}+\sum_{\ell=1}^{b-1} {q_{\ell,b}\over 2^\ell}\Biggr)= {1\over 2^z}~{\bf p}^T \cdot {\bf q}_{i;b}~,
\label{eq:deltra_comp}
\ee
where the prefactor $1/2^z$ decreases at each iteration by increasing $z$ (cf. Refs. \cite{Krak:2021,krakoff2022controlled} and Ref. \cite{Illa:2022jqb}). The  prefactor $1/2^z$ allows to control the refinement of our search,  increasing the precision of the gradient-descent method. It should be pointed out that by increasing the number of iterations, i.e.  zoom steps, one could decrease the initial number of bits $b$ in the expansion of the real vectors involved in the minimization, without worsening the accuracy of the algorithm (see also Refs. \cite{Krak:2021,krakoff2022controlled} for the symmetric GEVP). 

By expanding $f(C,{\bf w}^{z},\tilde\lambda)$  around ${\bf w}^{z-1}$ up to the second order, one gets (cf.  Eqs. \eqref{eq:objfc1} and \eqref{eq:sym_mat})
\be
\Delta f(C,{\bf w}^{z},\tilde\lambda) = 2{\bm \delta}^T(z)
{\cal S}(\tilde \lambda) {\bf w}^{z-1} +  {\bm \delta}^T(z)
{\cal S}(\tilde \lambda) {\bm \delta}(z)~. \nonu
\label{eq:hessian}
\ee
The term linear in ${\bm \delta}(z)$ can be recast in a QUBO form by i) recalling that $q_\ell= q^2_\ell$, being $q_\ell=0,1$, and ii) transforming the vector ${\cal S}(\tilde \lambda){\bf w}^{z-1}$ in a $(n\times b)\times (n\times b)$ diagonal matrix, where the vector components are the diagonal elements (see Ref. \cite{Krak:2021}). Then, the  OF suitable for the gradient-descent phase is obtained. Specifically, recalling that one  can divide or multiply by a constant an OF  without affecting the minimization procedure, one has
\be
\widehat f(C,{\bf \delta}(z),\tilde\lambda)={\bm \delta}^T(z)~ {\cal Q}(z,\tilde \lambda)~{\bm \delta}(z)
\label{eq:new_objf}\ee
where the square matrix $ {\cal Q}$ is given by
\be
\Bigl[{\cal Q}(z,\tilde \lambda)\Bigr]_{ij}= 2\delta_{ij}~\Bigl[{\cal S}(\tilde \lambda)~{\bf w}^{z-1}\Bigr]_i +
{\cal S}(\tilde \lambda)_{ij}~,
\label{eq:calQ}\ee
with the elements of the symmetric matrix ${\cal S}(\tilde \lambda)$ written in Eq. \eqref{eq:sym_mat}. The transformation to the binary expression of the vectors ${\bm \delta}(z)$ follows the  rule given in Eq. \eqref{eq:vect_bin}.

The  gradient-descent phase proceeds through iteration on $z$, with  $z=1,\dots,N_z$, and aims to minimize the OF in Eq. \eqref{eq:new_objf}, starting with the value of $\tilde \lambda
=\lambda^R({\bf w}_{{GP}})$
and the corresponding vector ${\bf w}_{{GP}}$. At each zoom step $z$,   an inner loop (labelled with $i$) is opened. Within this loop,  the \texttt{QA}, or the \texttt{SA}, returns  an ensemble of  $\alpha_i=1,2,\dots,N^{DP}_A$ qubits states,  from which we select the one with minimal energy, i.e.
\be
\widehat f^{DP}_{best;i}(z)= \min_{{\bm \delta}_{\alpha_i}(z)} 
\widehat f(C,{\bm \delta}_{\alpha_i}(z),\tilde\lambda=\lambda^R({\bf w}^z_{\alpha_i}))\,,
\ee
where $i$ is the inner-loop index for a fixed $z$ and ${\bf w}^z_{\alpha_i}= {\bm \delta}_{\alpha_i}(z)+{\bf w}^z_{best,{i-1}}$. 
For the value $\alpha_i=\alpha_{best}$ in correspondence of $\widehat f^{DP}_{best;i}(z)$, we calculate the values
 $\lambda^{R(I)}({\bf w}^z_{\alpha_{best}};i)$, that is  input for the next iteration starting with the new eigenvector ${\bf w}^z_{best,i}={\bf w}^z_{\alpha_{best},i}$.
When, for a fixed $z$, the condition $\widehat f^{DP}_{best;i+1}(z)\geq\widehat f^{DP}_{best;i}(z)$ is found, then the internal loop on $i$ is terminated and the value of $z$ updated, namely $z\rightarrow z+1$. The final best result obtained for a given  $z$ is used as starting point in the $z+1$ iteration.
The procedure is repeated until the desired precision is reached ($z^{max}$). 

Finally, one obtains the eigenvector of  the GEVP in Eq. \eqref{eq:gevp} by inverting the  relation in Eq. \eqref{eq:w_v}, i.e.
\be
{\bf v}_{best}= \Bigl [ L^T\Bigr]^{-1}{\bf w}_{best}^{z_{max}}
/\left|\left| \Bigl[ L^T\Bigr]^{-1}{\bf w}_{best}^{z_{max}}\right|\right |~,
\label{eq:v_best}\ee
where $z_{max}$ indicates the final run of the iterative gradient-descent method and
${\bf w}^{z_{max}}_{best}$ is the eigenvector corresponding to the final eigenvalue 
\be
\label{eq:lambdabest}
\lambda_{best}= \lambda^{R}({\bf w}_{best}^{z})\,.
\ee
It should be recalled that the Hessian is always positive, so that a minimization path is ensured. 
The   gradient-descent algorithm is sketched    in Tab. \ref{tab:desc_phase}.
\begin{table*}[htb]
\renewcommand{\arraystretch}{1.5}
\caption{The  gradient-descent phase search  of  the global minimum on both \texttt{SA} and \texttt{QA}. The gradient-descent phase is based on the Hessian of the OF in Eq. \eqref{eq:objf} (see text). The   same guess-phase number of bits  $b$  is  used. }
    \centering
    \begin{tabular}{c l}
   \toprule 
   ~& {\bf Algorithm  for the gradient-descent phase}\\
    \midrule 
        1:~ &  {\bf Use} $\tilde \lambda=\lambda^R({\bf w}_{{GP}})$,   from the guess phase\\ 
        2:~ & {\bf While} $1/2^z\ge \epsilon'$ ($z\ge 1$)  \\
        3:~ &  \hspace{0.5cm}Transform  the obj. function in Eq. \eqref{eq:new_objf} to QUBO form with $1/2^z$ in ${\bf p}$ \\
        4: ~& \hspace{0.5cm}{\bf While} $\widehat{f}_{best;z}^{DP}(i) < \widehat{f}_{best;z}^{DP}(i-1)$ for i =1,2,3,\dots  \\
        5:~& \hspace{1cm} Minimize the QUBO form in Eq. \eqref{eq:new_objf},  with $N^{DP}_A$ annealing cycles\\
        6:~ & \hspace{1cm} Search the minimal value among the  $N^{DP}_A$ energies returned by the annealer ($\alpha_{best}$)\\ 
        7:~&  \hspace{1cm} Replace $\tilde{\lambda}$ 
        with $\lambda^{R}_z({\bf{ w}}_{\alpha_{best}};i)$ \\
        8: &  \hspace{0.5cm}Pass to $z+1$ with $\tilde{\lambda}= \lambda^{R}_{z}({\bf w}_{\alpha_{best}};N_z)$ 
        \\
 \bottomrule
    \end{tabular}
  \label{tab:desc_phase}
\end{table*}

By repeating the entire minimization procedures, i.e guess plus gradient-descendent phases, given the  non-deterministic nature of the annealing process,    errors and fluctuations of the final results, $\{\lambda_{best},{\bf v}_{best}\}$, are generated. 
In order to study such uncertainties,  the  entire algorithm was run  by $N_{run}$ times, keeping fixed the initial matrices $A$ and $B$. This provides an estimate of the uncertainties associated to the annealing process, and how these uncertainties  propagate through the algorithm. We  performed this  statistical analysis using both the noisy annealer simulator, provided by D-Wave Systems software-package, with large $N_{run}$, and the \texttt{QA} with a substantially smaller $N_{run}$, given  the limitation  on running-time at our disposal.  In what follows,  the results obtained with the simulator are indicated by inserting \texttt{SA} at the beginning of the figure captions, while  the results obtained with the D-wave {\em Advantage System 4.1} are labelled by \texttt{QA}.

\section{Results}
\label{sect_results}
In order to study the reliability of the algorithm, we used the matrices $A$ and $B$ obtained from the discretization of the hBSE describing a  system of two massive scalars, with mass equal to $m$ and strong binding energy  $E_B/m=1.0$. They   interact through the ladder-exchange  of a massive scalar,  with  mass $\mu/m=0.15$   (see  Eq. \eqref{eq:kappa2} in Appendix \ref{app_hBSE}, and for more details  Ref. \cite{FSV2}).  In our  investigation we classically evaluate the $LDL^T$ factorization, exploiting the non singularity of the symmetric matrix $B$, and then we solve the standard eigenvalue problem given in Eq. \eqref{eq:objfc}. We leave to future works the generalized case. Let us recall that we focus only on the search of the largest real eigenpair. It should be recalled that, in our approach, the largest real eigenvalue   corresponds to the lowest coupling constant, which is able to  bind the  two massive scalars.
Finally, we give a first insight on  the scalability of the proposed algorithm by considering 
 matrix dimensions $n_{M}=4,8,12,16,24,32$.

 In an effort to  better optimize  the available quantum-computing resources,  a challenging issue  is to guide the  choice of the input parameters of the proposed algorithm. They are  given by: i)  the number of bits $b$, needed for the binary representation of the real vectors involved in the problem, and ii)   the number of annealing cycles, i.e. $N^{GP}_A$, for the guess phase, and $N^{DP}_A$, for the gradient-descent one. 
 This important  optimization task, which is time consuming,   was carried out by using the \texttt{SA}, and the  corresponding results  are discussed in the first part of this Section.
 In the second part, we illustrate the outcomes obtained by using the \texttt{QA}.
 
In particular, to perform the \texttt{SA} analysis, a  matrix with dimension $n_M=32$ was adopted and  the entire algorithm was run for  $N_{run}= 500$ times, in order to gather enough statistics and determine a reasonable set of  input parameters to be used with the \texttt{QA}. { For the remaining part of this work,  the  following symbols are adopted: i) $\bar\lambda_{best} (\bar\lambda_{best}^I)$ is  the average of  $\lambda^R (\lambda^I)$ obtained from Eq.~\eqref{eq:lambda_rn} (Eq.~\eqref{eq:lambda_in}) over the $N_{run}$ runs of the algorithm for $z^{max}$;  ii)  
$\overline{||{\bf v}_{true}-{\bf v}_{best}||}$  indicates the mean Euclidean distance, where ${\bf v}_{true}$ is the  real eigenvector  of the original GEVP, and ${\bf v}_{best}$ computed for $z^{max}$ is given in Eq. \eqref{eq:v_best}.}

\subsection{ Running the  guess-phase on the \texttt{SA}}
\label{sec:simulation:gp}
\begin{figure*}   
\includegraphics[width=\textwidth]{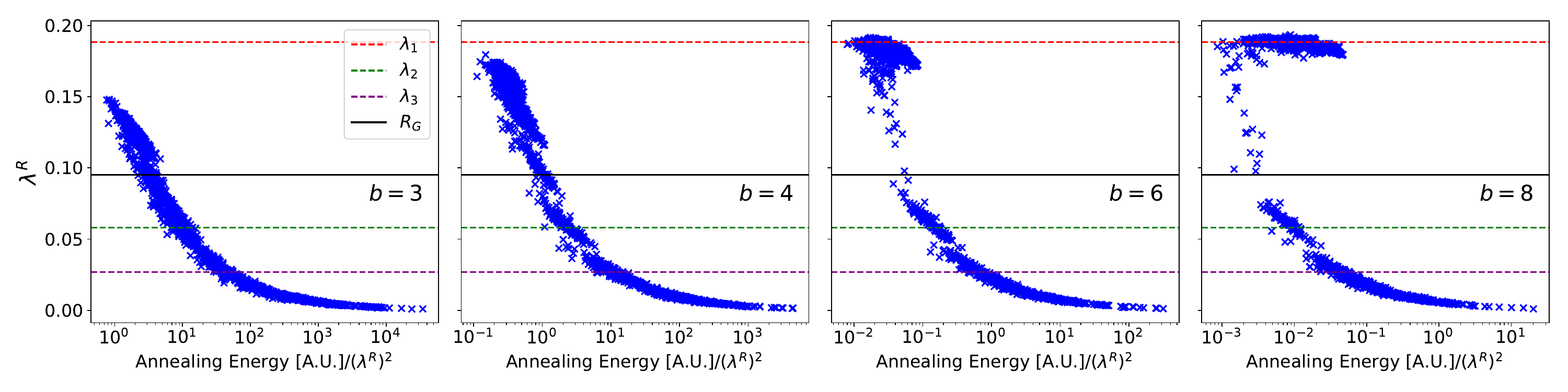}  \caption{\label{fig:E_vs_lambda}(Color online) [\texttt{SA}]   Scatter plot of the $N_A^{GP}=2000$ guess-phase values $\lambda^R_\alpha$, Eq.\eqref{eq:lambda_rn}, vs.  the corresponding annealer energies divided  by $[\lambda^R]^2$,
    for the bit  numbers $b=3,4,6,8$ used to get the QUBO matrix $P{\cal S}P^T$, Eq. \eqref{eq:objfc_bin}, with dimension $32\times b$.
    Dashed lines:  the three largest   real eigenvalues, $\lambda_1>\lambda_2>\lambda_3$ of the matrix $C$, Eq. \eqref{eq:C_def}.  { Solid line:  $|c_{11}-R_G(c_{11})|$ } with $R_G(c_{11})=\sum_{j\ne 1} |c_{1j}|$, i.e. the lower limit of the Gershgorin disc, with center $c_{11}=0.190$.}
\end{figure*}

Selecting a reliable initial guess in the first phase of the algorithm, sketched in Tab. \ref{tab:guess_phase}, is a crucial  step for obtaining an accurate result in the subsequent gradient-descent phase.
As already mentioned, all the real-valued eigenpairs are acceptable for the annealer, and therefore one has to guide the search toward the target solution, i.e. the real eigenpair with the largest real eigenvalue. In addition, one has to carefully consider another source of numerical challenges:    the (necessarily) finite number of bits $b$. In fact, one has  to assign $b$ for    representing  the components of the normalized vector entering the definition of the QUBO OF in Eq. \eqref{eq:objfc_bin}, and eventually construct the  matrix $P{\cal S}P^T$. This  obvious limitation can generate {\em spurious} minima,  and direct the path   toward   a wrong target. An example is shown in Fig.~\ref{fig:E_vs_lambda}, where the  case  of the three largest real  eigenvalues, $\lambda_1>\lambda_2>\lambda_3$, is studied. For bit numbers $b=3,4,6,8$ and running the entire algorithm only one time (i.e. $N_{run}=1$), the values of $\lambda^R ({\bf w}_\alpha)$,  Eq. \eqref{eq:lambda_rn},  obtained in the guess-phase from each of the $N^{GP}_A= 2000$ annealing cycles on the \texttt{SA}, are plotted  as a function of the corresponding annealer energies divided by $[\lambda^R]^2$ (this is suggested by the quadratic dependence of the OF on $\tilde \lambda$ and the purpose of  assigning a  normalization to different outcomes).
Notice that the annealer energies correspond to the values of Eq. \eqref{eq:objfc_bin} as returned by the annealer in the QUBO form i.e. with the binary vectors not normalized.

In Fig. \ref{fig:E_vs_lambda}, the  dashed horizontal lines represent the    three largest real eigenvalues (calculated classically) of the $32\times 32$ matrix 
$C$ in Eq. \eqref{eq:C_def}.  The solid line is the lower extremum of the Gershgorin disc, with center $c_{11}=0.190$. Such an extremum is  the lowest value admissible for $\lambda_1$. In the various panel, one observes   several regions where the results  appear to clusterize despite the initial guess  $\tilde \lambda=c_{11}$ is very close to the actual eigenvalue $\lambda_1=0.188225$.  Once a more detailed analysis is performed, it can be shown that   the clusterization is present already at
$b=3$,  although it is not observable in the figure.
 Moreover,  some clusters approach  the minima generated   by the other two eigenpairs, with eigenvalues $\lambda_2$ and $\lambda_3$, 
 while others are {\em spurious} minima, that are  generated by the limited precision of the assigned  binary representation, i.e. $3$ bits. By increasing $b$, the    clusters   migrate close to the actual eigenvalues, with smaller and smaller  energies (notice the different $x$-axis scale for the four panels). 
At the same time, a larger amount of points are approaching  $\lambda_1$, given the increasing precision,  although they  spread over a wider range of energy as expected. This is due to  the increase of states degeneration  when  the number of $b$ increases, since the returned bit strings  are  neither normalized nor orthogonal. 

Interestingly, $\lambda_1$ is approached from below by increasing $b$. This behavior can be understood considering the bias  due to  the spectrum of the matrix under scrutiny. In fact,  the decreasing ordering of the real eigenvalues, that follows   from the chosen  ordering of the matrix elements along the main diagonal, can  generate the observed   pattern, since the annealing process  should clearly spend a non negligible amount of time to explore the minima corresponding to the part of the spectrum below $\lambda_1$.  It turns out that the feature persists, even after running many times the code.


Increasing $b$ allows to refine the solution found by the annealer. On the other hand, a larger value of $b$ implies a larger  dimension of the matrix representing the OF. Unfortunately, such matrix dimension is constrained  by the actual topology of the qubits network during the process of minor-embedding (see below). This limits the precision of our solution in the initial guess phase.
Fortunately, as also discussed in Refs. \cite{Krak:2021,krakoff2022controlled},  a reasonable compromise  can be reached by properly increasing the number of  iterated annealing cycles and adopting a small number of bits, as investigated in the  following gradient-descent phase. 

\begin{figure*}   
\includegraphics[width=\textwidth]{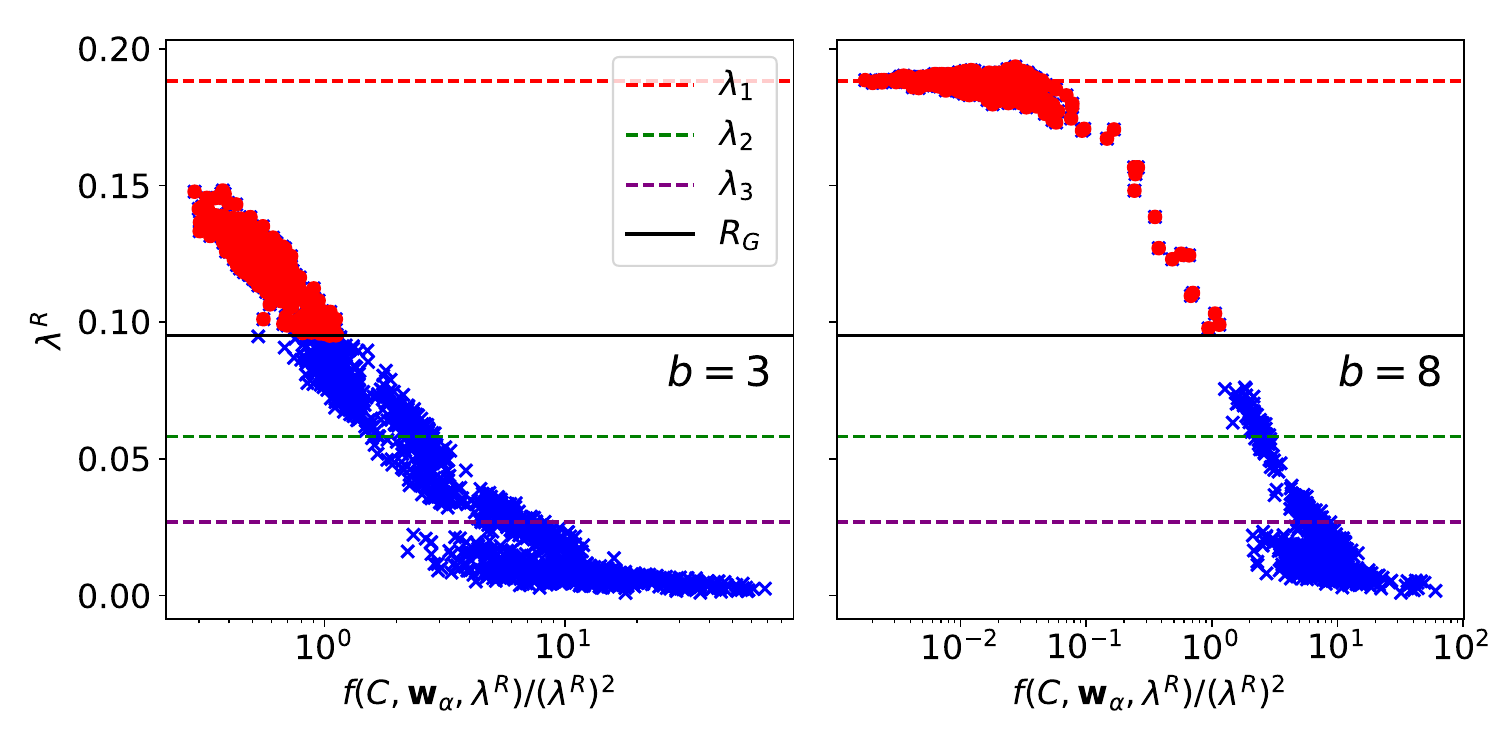} \caption{\label{fig:f_vs_lambda} (Color online). [\texttt{SA}] The same as the  panels $b=3$ and $b=8$, in Fig.~\ref{fig:E_vs_lambda}, but with the values of $f(C,{\bf w}_\alpha, \lambda^R)/(\lambda^R)^2$, Eq. \eqref{eq:objfc}, on the $x$-axis, instead of the annealing energies (see Eq. \eqref{eq:objfc_bin}).  Recall that here the vector ${\bf w}_\alpha$ is normalized. Red squares are the solutions selected after applying the $R_G(c_{11})$ cut, and blue crosses are the solutions outside the Gershgorin disc.}
\end{figure*}

In order to gain insights on the mapping of the OF in Eq. \eqref{eq:objfc} and the one given in terms of the binary strings, it is  useful   to show an analysis equivalent to the one in  Figs.~\ref{fig:E_vs_lambda}, but  focusing on the  values of the OF in Eq. \eqref{eq:objfc}, where normalized vectors are used.   In particular, the cases with $b=3$ and $b=8$ are shown   in Fig. \ref{fig:f_vs_lambda}, 
where the squares are the  solutions selected after
applying the Gershgorin criterium with $R_G(c_{11})$. The clusterization is more evident than in Fig.~\ref{fig:E_vs_lambda} and presents a different overall behavior compared to the results directly obtained from the annealing process. In particular,   the accuracy of the OF mapping onto the binary basis improves by increasing the number of bits, and the annealer spend more and more time close to the actual minima. 

Summarizing, two main consequences can be deduced from the calculations shown in Figs. \ref{fig:E_vs_lambda} and \ref{fig:f_vs_lambda}. First, in the  guess phase,  it is not possible to  select  the final  value of $\lambda^R$ only on the basis  of the annealer energies, as in the symmetric case (see, e.g., Refs. \cite{Krak:2021,krakoff2022controlled}), but we need the Gershgorin criterium, to overcome the issue of the quadratic dependence on $\tilde \lambda$ in the OF. 
Second, we   note that by exploiting  the mentioned quadratic dependence,  the annealer becomes a powerful tool to explore a wide portion of the spectrum of our eigenvalue problem. In principle, it is possible to identify also the other eigenpairs, by studying the clusterization of the returned energies. This appealing feature  will be investigated elsewhere.

\begin{figure}   
\includegraphics[width=\columnwidth]{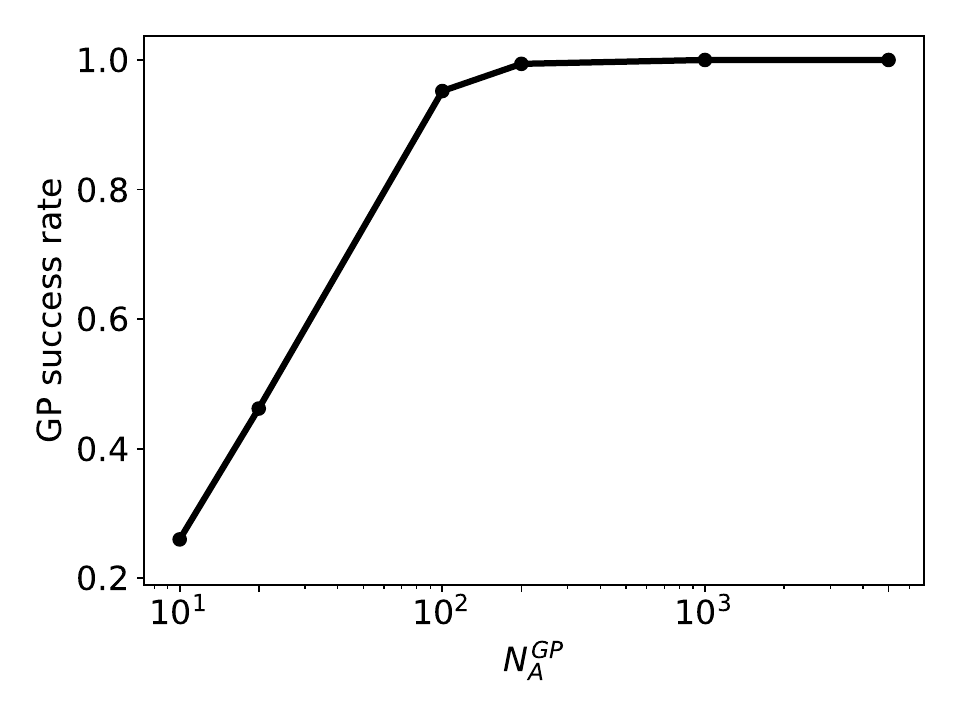}  
\caption{ \label{fig:sr_nagp}
[\texttt{SA}] Guess-phase success-rate vs. $N_A^{GP}$, for a $32\times 32$ matrix in Eq. \eqref{eq:objfc}. The solid line was obtained considering  $N_{run}=500$ independent samples with $b=3$ (see Tab. \ref{tab:guess_phase}  for the outline of the algorithm in the guess-phase).}
\end{figure}

In Fig. \ref{fig:sr_nagp}, the GP success-rate, i.e. the number of times  the algorithm is able to find a  solution falling inside the Gershgorin disc,  is shown as a function of the  GP annealing cycles, $N_A^{GP}$, for  $b=3$ and  running the code $N_{run}=500$ times for the $32\times 32$ matrix.  A number of $\sim 200$ annealing cycles seems to guarantee a $100\%$ probability of having at least one point within the Gershgorin disc, by using the \texttt{SA}. 

\begin{figure}
    \centering
    \includegraphics[width=\columnwidth]{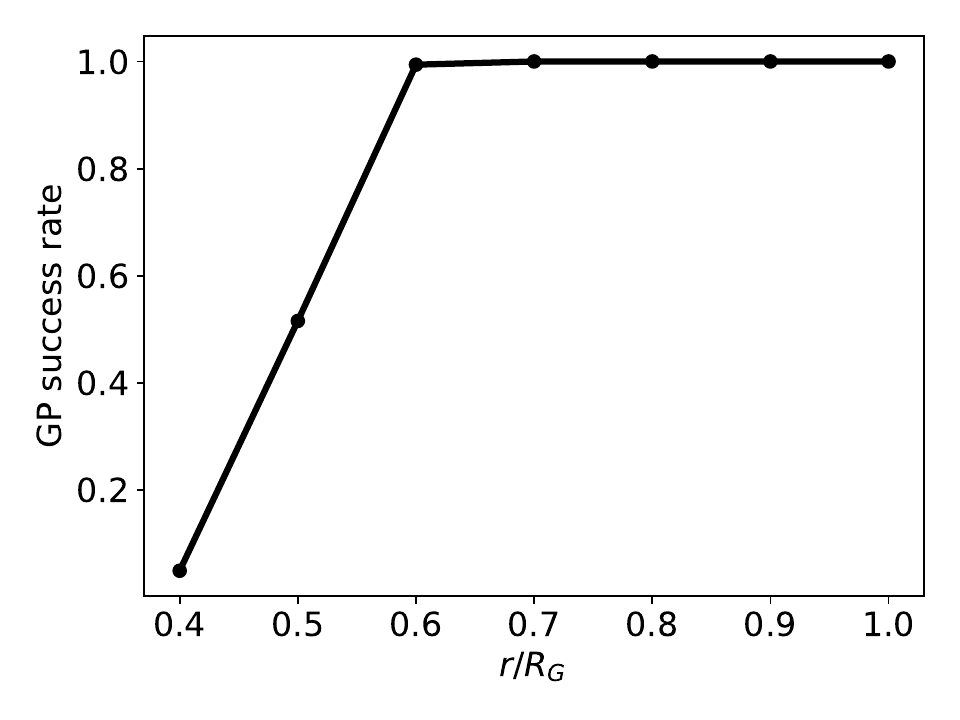}
    \caption{[\texttt{SA}] Guess-phase success-rate  vs. an arbitrary reduction of the Gershgorin radius (represented by the ratio $r/R_G$), for a $32\times 32$ matrix in Eq. \eqref{eq:objfc}.  The solid line was obtained by considering  $N_{run}=500$ independent samples, with $b=3$ and  $N_A^{GP}=200$ (see Tab. \ref{tab:guess_phase} for the outline of the algorithm in the guess phase).}
    \label{fig:lambda_vs_r}
\end{figure}

Since it is crucial to receive as reliable input as possible  from the guess phase, so that a better final outcome could be reached, we     attempted to  find  a tighter bound, if any,  for $\tilde \lambda$ to be used in the initial run of the gradient-descent phase.  In order to implement this,  we  arbitrarily reduced  the Gershgorin radius, $R_G(c_{11})$, to a chosen value  $r$.
 In Fig.~\ref{fig:lambda_vs_r}, the GP success-rate for the above $32\times 32$ matrix  is analyzed by i) taking $b=3$ and ii) running  $N_{run}=500$ times the GP algorithm with annealing cycles $N_A^{GP}=200$.  As expected,  the GP success-rate decreases  by reducing $r$, and    we were not able to find any solution below  $r/R_G<0.4$.
 In conclusion,  for the matrices considered  in this work,   the most reasonable and practical choice is $r=R_G(c_{11})$.

\subsection{Running  the  gradient-descent phase on the \texttt{SA}}
\label{sec:simulation:dm}

The main goal of the gradient-descent phase is  the improvement of the accuracy in determining the searched eigenpair. We pursued  this scope by decreasing the   prefactor $1/2^z$, with $z\ge 1$, in Eq. \eqref{eq:deltra_comp}.
In this subsection we focus on the impact of the input parameters by increasing this prefactor. Notice that for all the analysis  carried out in this section, we run   $N_{run}=500$ independent simulations.

\begin{figure}  
\includegraphics[width=\columnwidth]{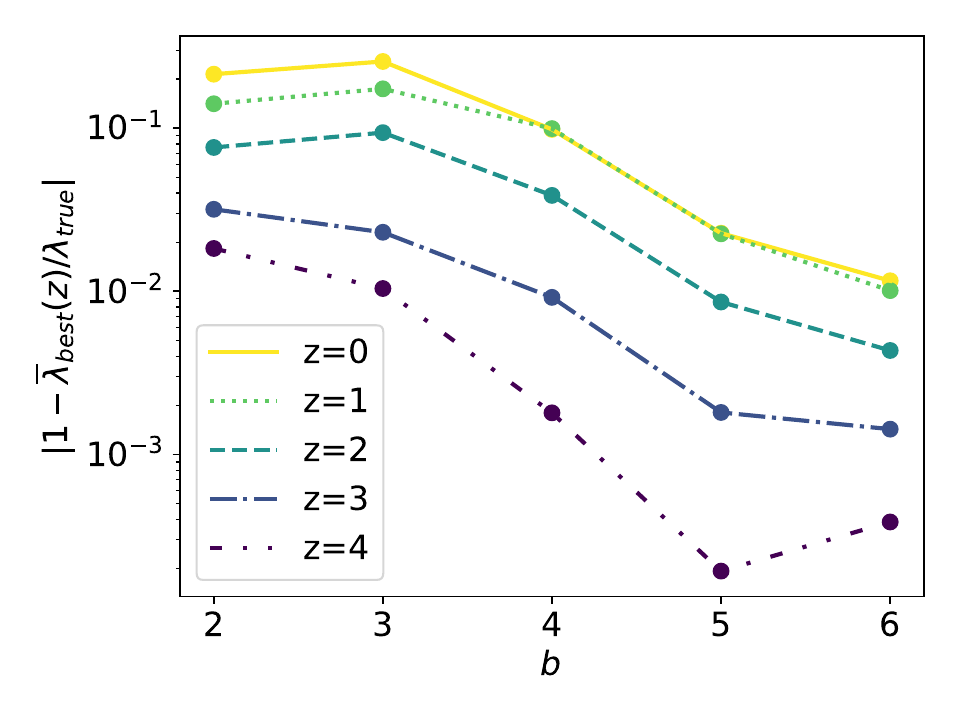}  \caption{\label{fig:lambda_vs_b}(Color online)[\texttt{SA}] Relative precision $|1-\bar\lambda_{best}(z)/\lambda_{true}|$, at fixed zoom step,  vs.  the number of bits $b$. In the calculations, we used $N_A^{GP}=200$,  $N_A^{DP}=20$. The dots represent the mean of the relative precision obtained from   $N_{run}=500$ independent samples. 
}
\end{figure}

 In Fig.~\ref{fig:lambda_vs_b},    the relative precision $|1-\overline{\lambda}_{best}(z)/\lambda_{true}|$ (cf. Eq. \eqref{eq:lambdabest}) dependence on the number of bits, $b\ge 2$, is presented for various zoom steps $z$ (in what follows, $z=0$ indicates the guess phase and the values $z\ge 1$ belong to the  gradient-descent phase). The dots in the figure are the mean, over $500$ independent runs, of the relative precision obtained by using 
 $N_A^{GP}=200$, and $N_A^{DP}=20$. An overall  decreasing pattern can be observed for increasing $b$ and fixed $z$, that  can be  largely ascribed  to a better identification of the minimum during  the guess phase, $z=0$ (see Section~\ref{sec:simulation:gp}).  
 Interestingly,  the numerical outcomes shown in Fig. \ref{fig:lambda_vs_b} suggest that  it is possible to establish a fair trade-off between the number of bits and the number of zoom steps in order to achieve increasingly accurate results, as already observed  for the symmetric GEVP (see, e.g., Refs. \cite{Krak:2021,krakoff2022controlled}). One should recall that the lower  the number of bits,  the lower  the dimension of the QUBO matrix, and thus  the less difficult  the eigenpair search becomes.

\begin{figure}   
\includegraphics[width=\columnwidth]{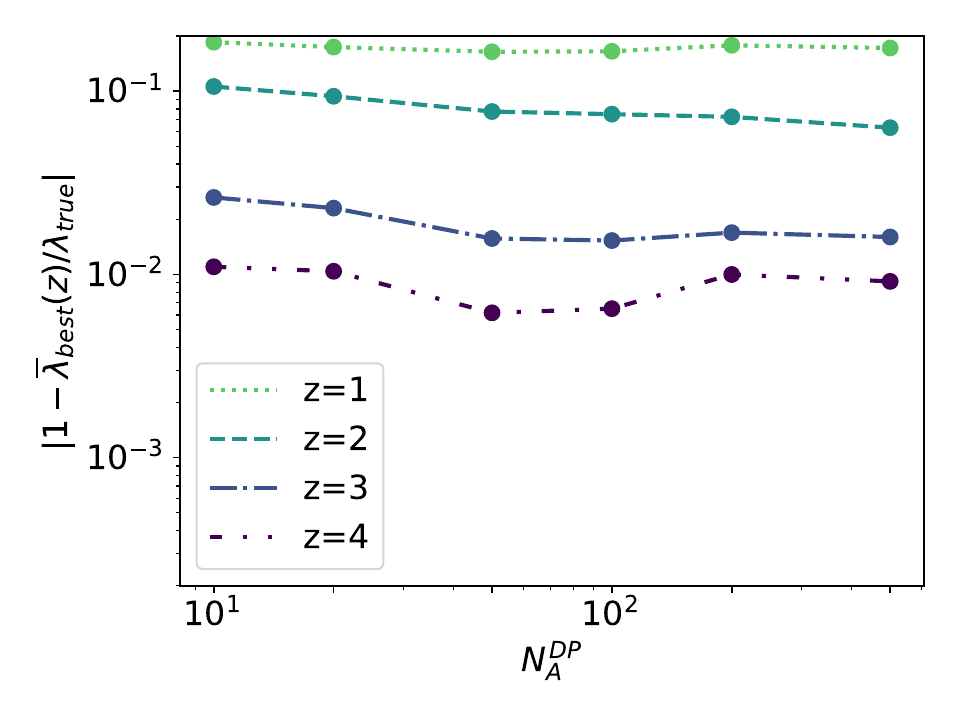}  \caption{\label{fig:lambda_vs_NADP}(Color online)[\texttt{SA}] The relative precision $|1-\bar\lambda_{best}(z)/\lambda_{true}|$, at fixed zoom step, vs.  $N_A^{DP}$, the number of annealing cycles  in the gradient-descent phase. In the calculations, $N_A^{GP}=200$ and   $b=3$  were used. The dots represent the mean of the relative precision performed over $N_{run}=500$.}
\end{figure}

In Fig.~\ref{fig:lambda_vs_NADP}, the relative precision  at  fixed zoom step was studied by changing the number of annealing cycles during the gradient-descent phase, $N_A^{DP}$, keeping fixed $N_A^{GP}=200$ and $b=3$. 
The figure shows that there  is basically no improvement by  increasing $N^{DP}_A$. In the gradient-descent phase for a given $z$, the algorithm requires a minimal number of annealing cycles to find the best minima for ${\bm \delta}(z)$. This is one of the advantage of using such an algorithm.

In conclusion, the campaign of numerical calculations performed on the \texttt{SA}  has suggested to adopt  the following input parameters for the  studies of our algorithm  on the \texttt{QA}:  $N^{GP}_A=200$,  $N^{DP}_A=20$, and $b=2$ or $b=3$, but with a suitable number of zoom steps. This preliminary analysis is critical to fine-tune the strategy when  the \texttt{QA} comes into play and    a  limited run-time is available.  It also sheds light on the different behavior of \texttt{QA} and \texttt{SA}, with obvious practical implications for the analysis of large matrices.
%
\subsection{ Analysis of the \texttt{QA} results}
We perform  the {\em experimental} study of the algorithm by using the  {\em Advantage system 4.1} quantum annealer, provided by D-Wave Systems. As a first step, we adopted $N^{GP}_A=200$ and $N^{DP}_A=20$ in order to study the whole set of matrices whose relevant eigenpairs we aim to determine. As  for the  number of bits, we used  $b=3$ for $n_M<16$ and $b=2$ for $n_M\ge 16$,   in order to avoid    limitations on the matrix dimensions that can be processed  by the \texttt{QA}.  In fact,  the current \texttt{QA} has a given upper-bound on the number of  logical qubits to be mapped onto sets of physical qubits, organized with   an assigned topology\footnote{This process is called  minor-embedding or simply embedding. Recall that      the matrix dimension in the QUBO problem  is  $n_M\times b$, and therefore  $n_M\times b$ logical qubits are fully connected with weights given by the values of the matrix elements. For the {\em Advantage 4.1 system} \cite{mcgeoch2021advantage,mcgeoch2022advantage},  based on the Pegasus topology of the physical qubits,  a maximal size of a fully-connected set of logical qubits (clique) is $177$, but it can be reached under very peculiar conditions, not fully achievable in our actual problem, if accurate results  have to be obtained. }

In Tab. \ref{tab:results_qa} we show the summary of the \texttt{QA} results for the problem in Eq. \eqref{eq:gevp}, with   
 dimensions $n_M=4,8,12,16,24,32$. As target we selected a nominal relative precision on the eigenvalue of $\epsilon'=10^{-3}$ that in the algorithm corresponded to a value of $z=9$. 
 
In the fifth column, we present the mean value  $\bar\lambda_{best}$ obtained running the algorithm $N_{run}$  times. The agreement between  the {\em true}  eigenvalue, $\lambda_{true}$ in the fourth column, and  the mean value $\bar \lambda_{best}$ is smaller than the nominal precision requested  $\epsilon'=10^{-3}$. More importantly,   all the results obtained in our runs are better than the requested precision. This can be observed by comparing the $68\%$ confidence  interval over the $N_{run}$ runs with the nominal precision $\epsilon'=10^{-3}$.

In Tab. \ref{tab:results_qa} we also show: i)  the ratio $\bar\lambda^I_{best}/\bar\lambda_{best}$, and ii)  the mean Euclidean distance 
$\overline{||{\bf v}_{true}-{\bf v}_{best}||}$. It is worth noting that the ratio and the mean Euclidean distance have almost the same magnitude,  
and  slightly increases  with the matrix dimension $n_M$. In view of this,   a remarkable degree of reliability  of  the proposed algorithm can be inferred.

\begin{table*}[htb]
\renewcommand{\arraystretch}{2.0}
    \centering
    \begin{tabular}{cccclll}
    \toprule 
       $n_M$ & $b$ & $N_{run}$ & $\lambda_{true}$&$~~\bar\lambda_{best}$  & $\bar\lambda^I_{best}/\bar\lambda_{best}$ & $\overline{||{\bf v}_{true}-{\bf v}_{best}||}$  \\
       \midrule 
       4  & 3 & 80 & $0.188026$& $ \newss 0.188012^{+ 1 \cdot 10^{-5}}_{- 7 \cdot 10^{-6}}$ &  $  0.00024^{+ 2 \cdot 10^{-5}}_{- 5 \cdot 10^{-6}}$ & $   0.00024^{+ 3 \cdot 10^{-5}}_{- 2 \cdot 10^{-5}}$  \\
       8  & 3 & 80 & $0.188204$& $0.18820^{+ 2 \cdot 10^{-5}}_{- 2 \cdot 10^{-5}}$ & $0.0003^{+ 1 \cdot 10^{-4}}_{-1 \cdot 10^{-4}}$& $0.0003^{+1 \cdot 10^{-4}}_{-1 \cdot 10^{-4}}$ \\
       12 & 3 &80 & $0.188203$& $0.18821^{+ 2 \cdot 10^{-5}}_{- 2 \cdot 10^{-5}}$ & $ 0.0005 ^{+ 1 \cdot 10^{-4}}_{- 1 \cdot 10^{-4}}$ & $0.0006^{+2 \cdot 10^{-4}}_{-1 \cdot 10^{-4}}$ \\
        16 & 2 & 80 & $0.188203$& $0.18820^{+ 4 \cdot 10^{-5}}_{- 3 \cdot 10^{-5}}$ & $0.0009^{+ 1 \cdot 10^{-4}}_{- 1 \cdot 10^{-4}}$ & $0.0011^{+ 2 \cdot 10^{-4}}_{- 2 \cdot 10^{-4}}$  \\
       24 & 2 & 80 & $0.188225$ & $0.18822^{+ 5 \cdot 10^{-5}}_{- 4 \cdot 10^{-5}}$ &  $0.0013^{+ 1 \cdot 10^{-4}}_{- 3 \cdot 10^{-4}}$ & $0.0015^{+ 3 \cdot 10^{-4}}_{- 2 \cdot 10^{-4}}$\\
       32 & 2 & 200 & $0.188225$& $ 0.18823^{+ 4 \cdot 10^{-5}}_{- 3 \cdot 10^{-5}}$ & $0.0016^{+ 2 \cdot 10^{-4}}_{- 2 \cdot 10^{-4}}$  &$0.0018^{+4 \cdot 10^-4}_{-3 \cdot 10^-4}$\\ 
       \bottomrule
    \end{tabular}
    \caption{[\texttt{QA}] Results obtained by using the  {\em Advantage 4.1} quantum annealer, provided by D-Wave Systems. In the Table, $n_M$ is the dimension of the matrices  $A$, $B$ and   $C$ involved in the calculations (see Eqs.  \eqref{eq:gevp} and \eqref{eq:C_def}, respectively); $b$  is the number of bits adopted for the matrix representation in the QUBO problem (see Eq. \eqref{eq:objfc_bin}); $N_{run}$ is the number of runs of the entire algorithm (guess phase + gradient-descent phase); $\bar\lambda_{best}$ is  the  average of $\lambda_{best}(z_{fin})=\lambda^{R}({\bf w}_{\alpha_{fin}}^{z_{fin}})$ (cf. Eq.~\eqref{eq:lambda_rn} and the definition below Eq. \eqref{eq:v_best}), over the outcomes returned  after running  the entire algorithm $N_{run}$  times; 
  in $\bar\lambda^I_{best}/ \bar\lambda_{best}$  the numerator $\bar \lambda^I_{best}$   is   obtained as $\bar\lambda_{best}$, but following Eq. \eqref{eq:lambda_in}; $\overline{||{\bf v}_{true}-{\bf v}_{best}||}$ is the mean Euclidean distance between the    true eigenvector and the one  from the \texttt{QA} (recall that the vectors ${\bf v}_i$ are normalized). In addition to the mean values,    the $68\%$ confidence interval is shown. All the results are obtained using $z=9$ zoom steps, corresponding to an upper-bound for the relative precision equal to $\epsilon'=10^{-3}$.}
    \label{tab:results_qa}
\end{table*}

\begin{figure} [htb]  
\includegraphics[width=\columnwidth]{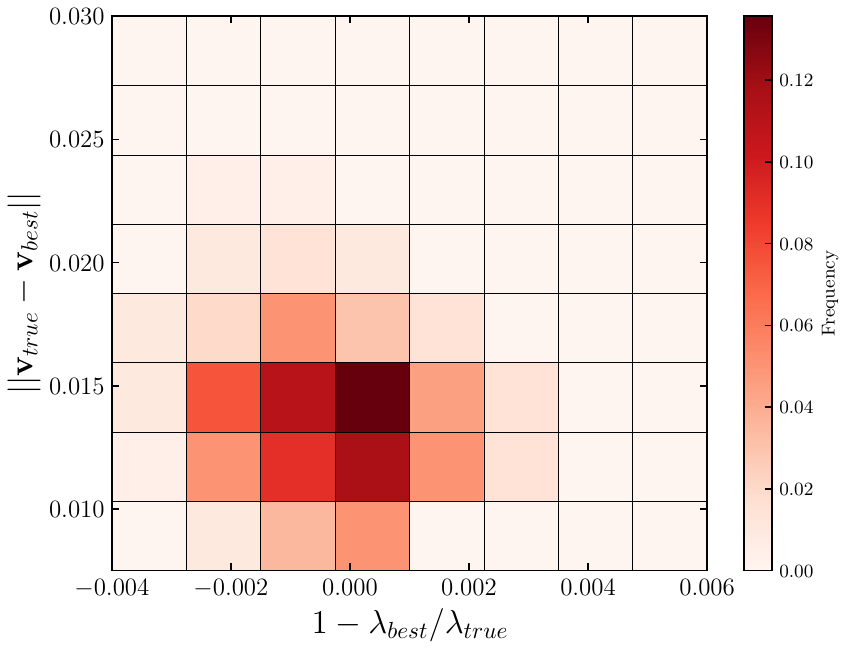}  
\includegraphics[width=\columnwidth]{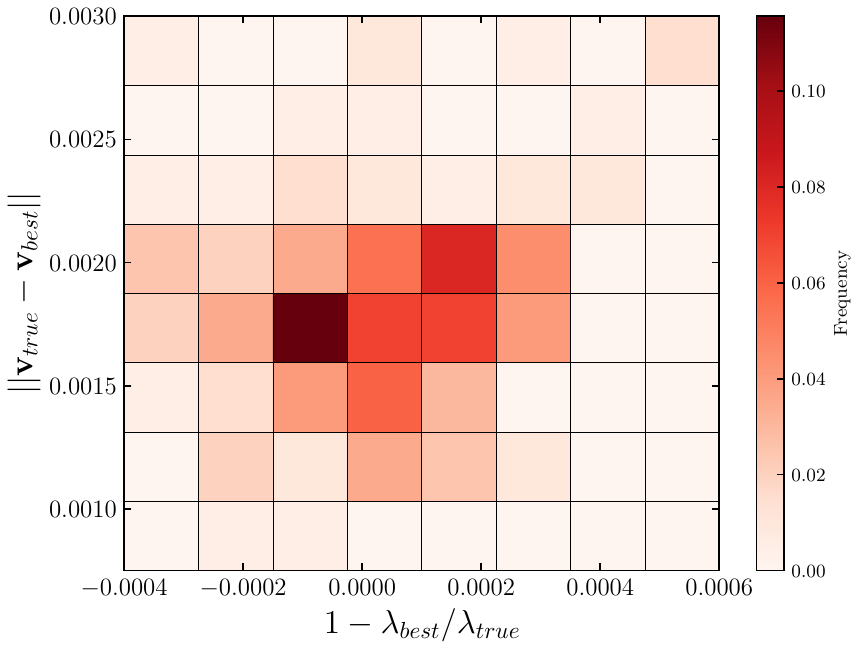}  
\caption{ \label{fig:Eigenvalue_vs_Eigenvector_32}
(Color online) [\texttt{QA}] Two dimensional distribution of the results for the  $n_M=32$ case,  after running  $N_{run} = 200$ times the entire algorithm as a function of the Euclidean distance $||{\bf v}_{true}-{\bf v}_{best}||$  vs.  the precision $(1-\lambda_{best}/\lambda_{true})$. Upper panel: $z=6$ results. Lower panel: $z=9$ results.  }
\end{figure}

In Fig.~\ref{fig:Eigenvalue_vs_Eigenvector_32} we plot the two dimensional distribution  of the results obtained running the code for $N_{run}=200$ in the $n_{M}=32$ case for $z=6$ and $z=9$ as function of $(1-\lambda_{best}/\lambda_{true})$, and the  Euclidean distance $|| {\bf v}_{true}-{\bf v}_{best}||$,  where $\lambda_{best}$ and ${\bf v}_{best}$ are the solution found by the algorithm (notice the difference with $\overline \lambda_{best}$). It is worth noticing that the relevant area substantially shrinks when passing from the distribution of the results for $z=6$  to $z=9$, as shown by the changes in the $x-y$ scales, but  it is not possible to identify a direct correlation between  the two quantities. 
However, the results seems to accumulate on a circular region that does not change shape increasing $z$ indicating that the correlation does not depend on the specific zoom phase.
We observed also that the projection on the $x$ axis, the distribution of $(1-\lambda_{best}/\lambda_{true})$,   approaches a Gaussian distribution  centered in zero  with a smaller and smaller width as  $z$ increases, while the projection on the $y$ axis, the Euclidean distance,  starts from zero and extend to positive values, still with a Gaussian fall-off.
 Plainly, these distributions are very important for a better understanding of the quantum-hardware behavior and should be carefully investigated   in a future work, when a much longer \texttt{QA} running-time  will be available for exploring  larger $N_{run}$ and lower $\epsilon '$.


\begin{figure} [htb]  
\includegraphics[width=\columnwidth]{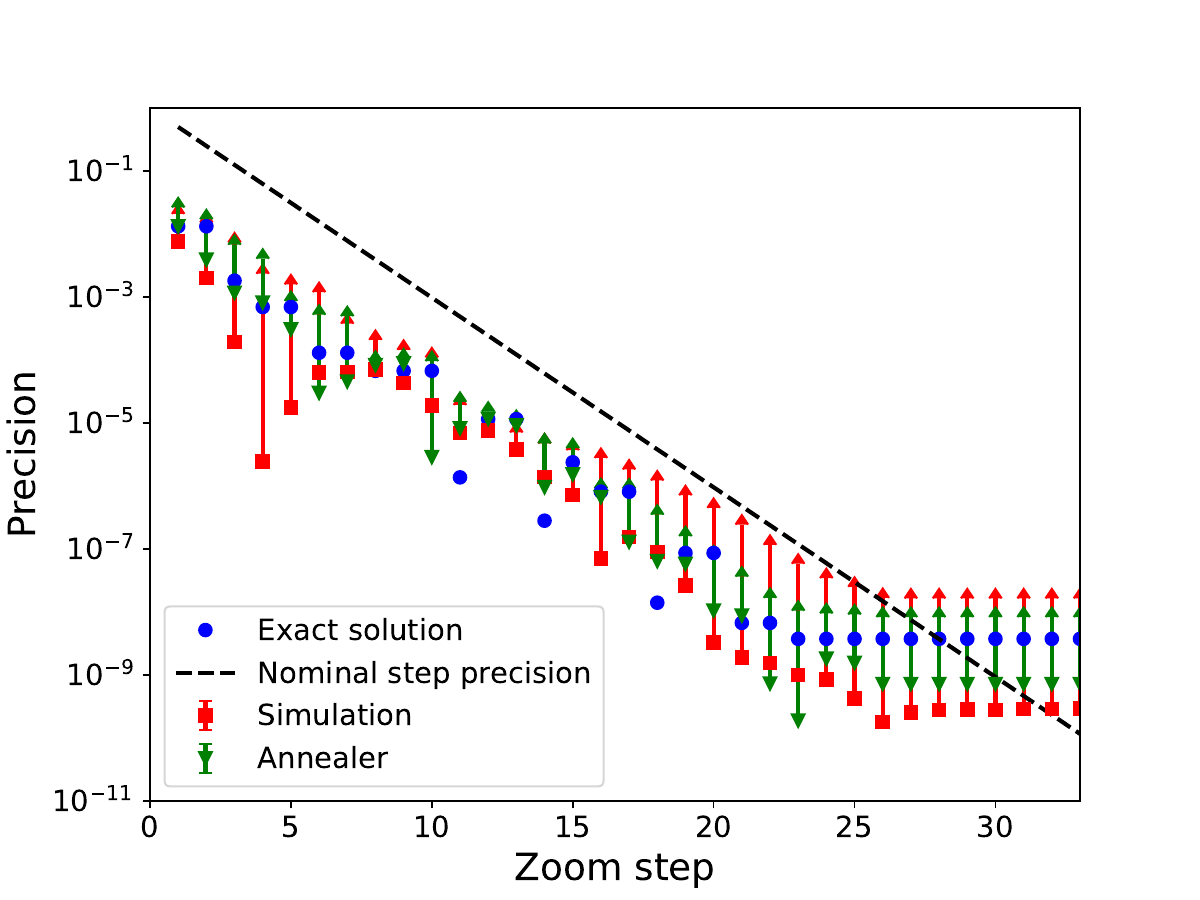}  
\caption{ \label{fig:precision_lambda_4x4}
(Color online) [\texttt{SA - QA}] Mean relative precision,  $|1 -\bar \lambda_{best}(z)/\lambda_{true}|$, vs.  the zoom step in the gradient-descent phase, for  $n_M=4$ case with $b=3$. The dashed line is the   nominal precision at each zoom step, equal to $1/2^z$. Red squares and arrows represent the mean and  the worst result respectively, obtained by averaging over $N_{run}=1000$ runs on the \texttt{SA}. Green triangles and arrows are the same as before obtained with the \texttt{QA}, but averaging   over $N_{run}=10$ runs. Blue dots are the results obtained using the exact solver for the QUBO problem at each zoom step. } 
\end{figure}

\begin{figure} [htb]  
\includegraphics[width=\columnwidth]{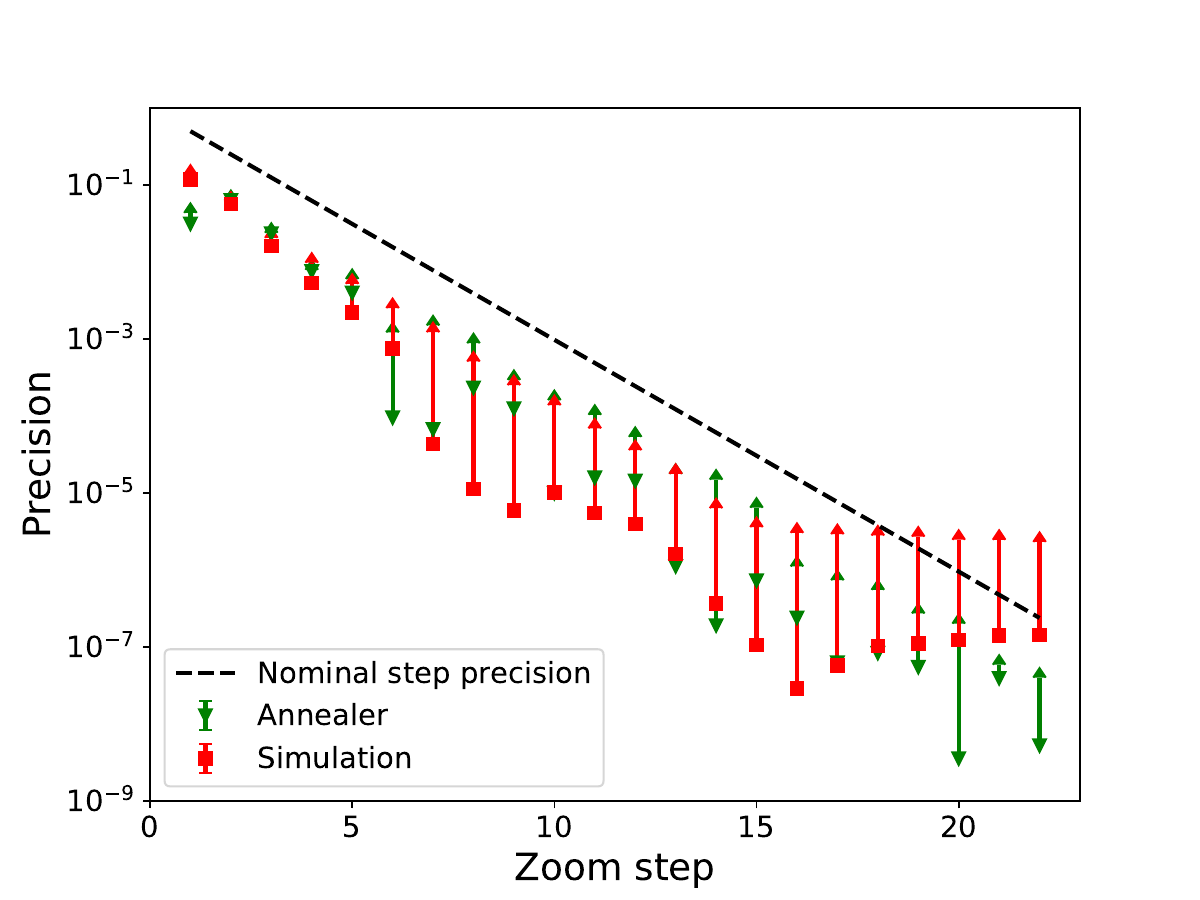}  
\caption{ \label{fig:precision_lambda_32x32}
(Color online) [\texttt{SA - QA}] The same as Fig.~\ref{fig:precision_lambda_4x4}, but  for  $n_M=32$ and $b=2$. The \texttt{QA} results   are obtained by averaging over   $N_{run}=8$ runs (see text). } 
\end{figure}

\begin{figure} [htb]  
\includegraphics[width=\columnwidth]{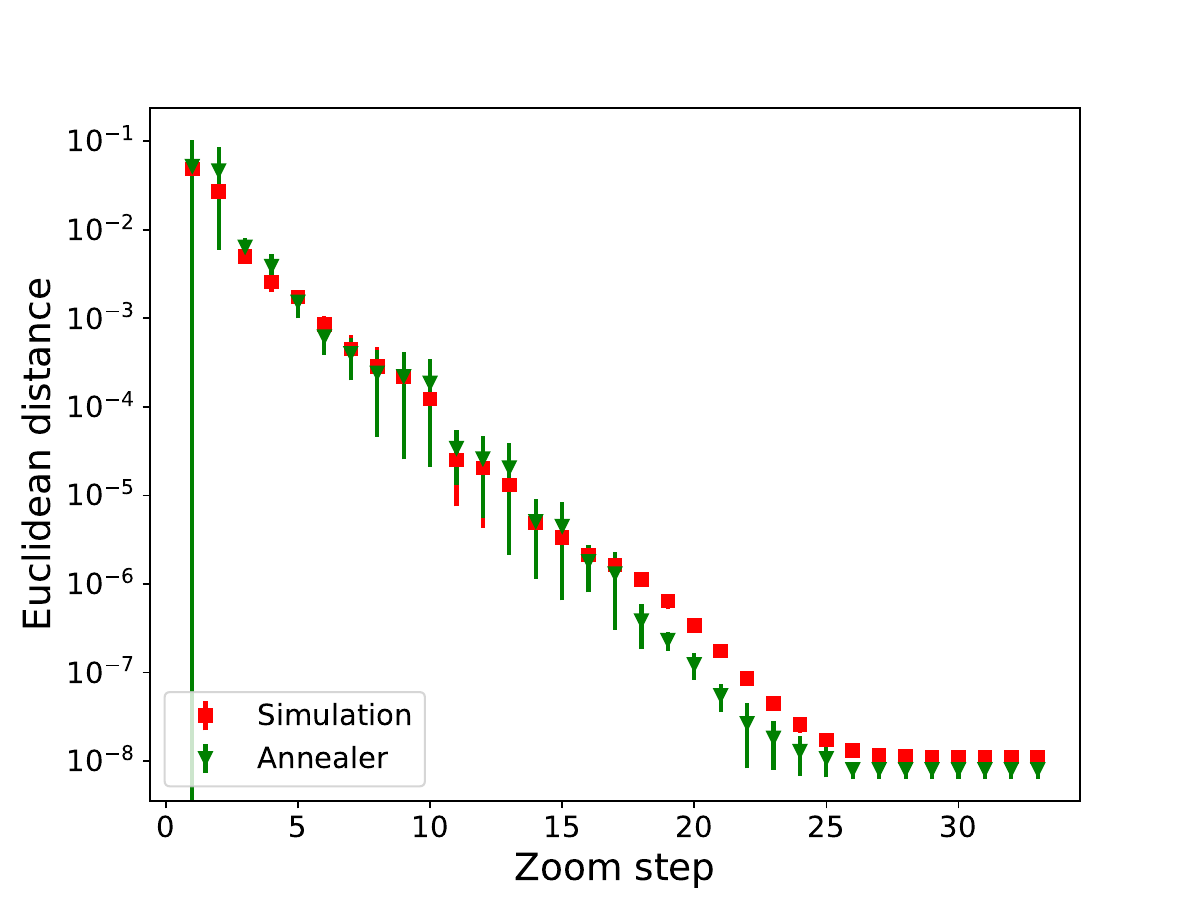}  
\caption{ \label{fig:precision_v_4x4}
(Color online) [\texttt{SA - QA}] Mean Euclidean distance between the  target eigenvector and the true one, at each zoom step, for the $n_M=4$ case with $b=3$. Green triangles: results from  the \texttt{QA}. Red squares: outcomes from the \texttt{SA}. Error bars represent the $99\%$ confidence level over 10 (1000) runs on the \texttt{QA} (\texttt{SA}).} 
\end{figure}

\begin{figure} [htb]  
\includegraphics[width=\columnwidth]{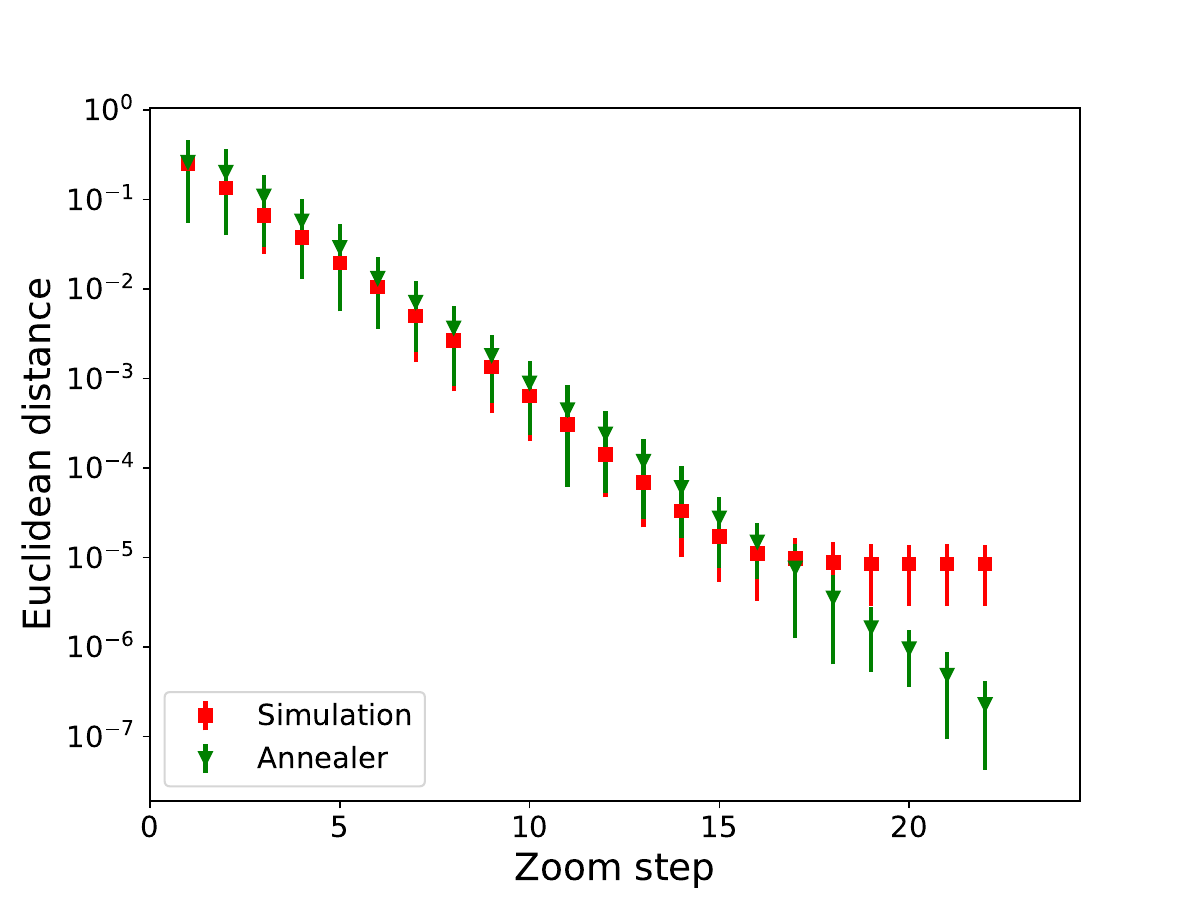}  
\caption{ \label{fig:precision_v_32x32}
(Color online) [\texttt{SA - QA}] The same as Fig.~\ref{fig:precision_v_4x4}, but  for  $n_M=32$ and $b=2$. The \texttt{QA} results   are averaged over  $N_{run}=8$ runs.} 
\end{figure}

Finally, an interesting comparison between the outcomes obtained  for matrix dimensions $n_M=4$ and $32$ from  \texttt{SA} and \texttt{QA}, is shown in Figs. \ref{fig:precision_lambda_4x4}, \ref{fig:precision_lambda_32x32}, and \ref{fig:precision_v_4x4}, \ref{fig:precision_v_32x32}, for both  the  eigenvalue  and   the eigenvector, respectively.  For  matrix  dimensions $n_M=4$, the analysis has been carried out  by using $b=3$ and pushing  the number of zoom step beyond   $z=30$   (recall that in  the calculations shown in previous figures the maximal value is $z=9$, leading to a nominal precision $\epsilon'=10^{-3}$).
For the case $n_M=32$, the number of bits was reduced to   $b=2$, as already explained, and  the zoom steps exceeded $z=20$. 

In  Figs. \ref{fig:precision_lambda_4x4} and \ref{fig:precision_lambda_32x32}, the mean relative precision $|1 -\bar \lambda_{best}(z)/\lambda_{true}|$ for fixed $z$  is shown for $n_M=4$ and $n_M=32$, respectively. Red squares and arrows represent the mean and  the worst result  obtained over $N_{run}=1000$ simulations, respectively. Green triangles and arrows are the same as before, but running  the entire algorithm on  \texttt{QA}  by $N_{run}=10$ times for $n_M=4$ and by $N_{run}=8$ times  for $n_M=32$. We reduced the number of runs used for this study compared to the one in Table~\ref{tab:results_qa}, because of the larger running-time requested to go beyond $z=9$. Finally, blue dots are the results obtained using the exact, classical   solver for the QUBO problem at each zoom step. These blue dots are not available for $n_M=32$, since it should be necessary to poll $2^{64}$ states and infer the suitable distributions (i.e. the well-known Feynman's argument in favor of a quantum computer).

As clearly seen in Figs. \ref{fig:precision_lambda_4x4} and \ref{fig:precision_lambda_32x32},  the actual accuracy  improves substantially as  $z$ increases for both \texttt{SA} and \texttt{QA}. In general, it remains below the nominal  precision  $1/2^z$, given by the dashed lines. For $n_M=4$, the results obtained with  \texttt{SA} and \texttt{QA} are below $1/2^z$  up to $z=25$, reaching a plateau  with a rough value of $\sim 10^{-8}$. For $n_M=32$,  the \texttt{SA} reach a plateau roughly  around $\sim 10^{-6}$ while the \texttt{QA} shows a behavior close to the one seen for $n_M=4$. From the comparison between $n_M=4$ and $n_M=32$ \texttt{SA} results, one could ascribe the different behavior  to a lower \texttt{SA}'s performance  in managing large matrices, rather than an issue with the algorithm. In fact, the results obtained with the \texttt{QA} just follows  the expected behavior.  One could understand heuristically the presence of a plateau by considering that the actual zero of the OF  $f(C,{\bf w},\tilde \lambda)$ is   numerically  about $10^{-16}$, and therefore a precision for  $\tilde \lambda$ of the order of  $\sqrt{f} $, i.e.  $\sim 10^{-8}$, is expected. Properly rewriting Eq. \eqref{eq:objfc}, one gets
\be
{f(C,{\bf w},\tilde \lambda)\over |{\bf w}|^2}=\Bigl[\tilde \lambda -\lambda^R({\bf w})\Bigr]^2+\Bigl[\lambda^I({\bf w})\Bigr]^2~,
\ee
 that for real $\tilde \lambda$ leads to the above estimate.

  The Euclidean distance between the target eigenvector and the results obtained, at each zoom step, with the \texttt{SA} and the \texttt{QA}, is shown in Figs.~\ref{fig:precision_v_4x4} and~\ref{fig:precision_v_32x32}, for  $n_M=4$ and $n_M=32$, respectively. The symbols are the same as in Figs. \ref{fig:precision_lambda_4x4} and \ref{fig:precision_lambda_32x32}, while the error bars represent the $99.9\%$ confidence level. As expected, the Euclidean distance   decreases by increasing $z$, until it reaches a plateau following a  pattern similar to the eigenvalue relative precision. Again, for $n_M=32$, the \texttt{QA} performs  better than the \texttt{SA}.

  Summarizing, the proposed algorithm opens an actual window on the study of the spectrum of non symmetric matrices by using a \texttt{QA}, { that nicely seems outperforms the software providing the simulated annealing. Clearly, the main issue is given by the size of the matrices to be used, but this will be surely overcome  by  the planned improvements of the QPU topology, so that  larger and larger matrices will be embedded.}


\subsection{Scalability  on the \texttt{QA}}

A first study of  the scalability of the proposed hybrid algorithm running on the \texttt{QA} has been carried out by focusing on the total time spent in the  annealing cycles.  In our case, the total annealing-time is theoretically given by  $3\times N^{GP}_A\times  t + z^{max}\times  i^{max}\times (N^{DP}_A \times t)$ (cf. Tabs. \ref{tab:guess_phase} and \ref{tab:desc_phase}) where $t\sim20\mu$s is the annealing time selected,  $z^{max}$ is the maximal value of zoom steps adopted in the gradient-descent phase, e.g. for the results shown in the third column of Tab. \ref{tab:results_qa} one has  $z^{max}=9$. Moreover, $i^{max}$  is the maximal number of search cycles for a given $z$ (see step 4 in Tab. \ref{tab:desc_phase}).  In  Tab.~\ref{tab:results_scala},  $T [m s]$ is the total annealing-time averaged on $N_{run}$ runs of the two-phase algorithm. One notices that a slightly increasing behavior is present when  the matrix dimension, $n_M \times b$, increases, at fixed $b$. But,  further studies with larger matrices  will be necessary for fully assessing a linear increasing of  the total annealing time on the \texttt{QA}.

 Another parameter to be considered when studying the scalability of the algorithm is
the number of physical qubits, $N_{qubits}$, used by the annealer to map the logical qubits of the QUBO problem on the topology of the hardware (Pegasus in this case), through the embedding process.
In general, since  each physical qubit is not connected to all the other, the mapping is  not a one-to-one correspondence, but it is a  one-to-many, i.e. one logical qubit is represented by a set of physical qubits, all in principle  forced to have the same value. 
This formal step is elaborated by  the proprietary software D-Wave {\em  Ocean}, before a  problem is submitted to the quantum annealer. 
The 
  software is heuristic and a different embedding, with a different number of physical qubits, is used  at each QA run. However,
 the exact number of physical qubits  adopted for mapping  the QUBO square matrix is provided by the software, so that   the distribution of $N_{qubits}$ over different runs can be eventually obtained.
The mean value and the $68\%$ confidence level of the number of this distribution is then showed in the last column of Table~\ref{tab:results_scala}.  It should be noted that the number of involved physical qubits is roughly given by the matrix dimension times the so-called chain length, i.e. the number of qubits needed for representing a single logical qubit. In our analysis, the chain length linearly grows with the matrices dimension as shown in Ref. \cite{mcgeoch2021advantage}.\footnote{ In Ref. \cite{mcgeoch2021advantage}, for {\em Advantage 4.1}, it is shown that one  can reach  at most   $N_{clique}=177$ fully-connected logical bits,  with a chain-length equal to 17, while for  $N_{clique}=647$ the chain length is about 7, not too far from what we found. It should be recalled the critical role of the accuracy one aims to, as well as the condition number of the involved matrices.  In our case, the matrix elements,  being  the outcomes of an actual physical problem, are within an interval several  orders of magnitude wide.} 
 The overall memory required by the algorithm grows therefore quadratically with the dimension of the original matrix.

\begin{table}[htb]
\renewcommand{\arraystretch}{2.0}
    \centering
    \begin{tabular}{cccc}
    \toprule
       $n_M\times b$ &  $N_{run}$ & $T [m s]$ & $N_{qubits}$\\
       \midrule
       12(4x3)  & 80  & $\newss 18.5^{+0.9}_{-0.7}$ & $24.1^{+0.9}_{-1.1}$ \\
       24(8x3)  &  80  & $21.7^{+1.3}_{-0.9}$ & $81.3^{+2.1}_{-2.3}$\\
       32(16x2) & 80   & $23.0^{+1.4}_{-1.0}$ & $140.9^{+1.9}_{-1.9}$\\
       36(12x3) & 80   & $23.0^{+1.4}_{-1.3}$ & $177.6^{+6.4}_{-6.6}$\\
       48(24x2) &  80  & $23.0^{+1.0}_{-1.0}$ & $306.6^{+11.8}_{-15.6}$ \\  
       64(32x2) &  200 & $23.6^{+1.2}_{-1.2}$ & $529.8^{+30.6}_{-33.8}$\\ 
       \bottomrule
    \end{tabular}
    \caption{[\texttt{QA}]  Total annealing time, $T[m s]$, and total number of physical qubits $N_{qbits}$, averaged on   $N_{run}$ runs, for the results shown in Tab. 
    \ref{tab:results_qa}.  }
    \label{tab:results_scala}
\end{table}

\section{Conclusions and Perspectives}
\label{sect_conclusions}
A hybrid algorithm, suitable for a quantum annealer, was implemented to evaluate the largest real eigenvalue and corresponding eigenvector of a generalized eigenvalue problem involving a non symmetric matrix. This numerical problem stems from the discretization of the homogeneous Bethe-Salpeter equation describing a bound state of two massive scalars, that interact  by exchanging  a massive scalar (see, e.g., Ref. \cite{FSV2}). In the current initial stage of the study, the non singularity of the symmetric matrix was exploited, so that a classical $LDL^T$ factorization of the  symmetric matrix was used in order to deal with  a simpler  QUBO problem. 

 The numerical results were  obtained by running our two-phase alogorithm   both on an {\em Advantage 4.1} quantum annealer, provided by the D-Wave Systems, and a simulator, based on the proprietary software {\em Ocean}.  We first tested our algorithm, based on an OF suggested in Ref.~\cite{alliney1992variational} and supplemented by the result of the valuable Gershgorin circle theorem \cite{varga2010gervsgorin}, on  the \texttt{SA}. This first investigation led to establish
 a practical set of input parameters, given by: i) the number of bits for expressing the real components of the involved vectors in the binary basis, and  ii) the number of annealing cycles  in both the guess phase and the gradient-descent one. Notably,  the studies carried out on the \texttt{SA} have confirmed that the trade-off between the bit number and the zoom-step number, that controls the nominal precision, is favorable as already found in the symmetric case \cite{Krak:2021,krakoff2022controlled}. Then, a minimal bit number can be chosen, so that the matrix dimension in the QUBO problem does not exceed the current limitations of the quantum hardware. After strengthening our numerical experience on the \texttt{SA}, we performed a numerical campaign by running $N_{run}$ times our algorithm  on the D-Wave \texttt{QA}, obtaining very encouraging results. We successfully approached the target eigenpair, extending our studies up to a matrix dimension $n_M=32$, with a corresponding QUBO-matrix dimension equal to $64$. 
 We have shown that the algorithm used in combination with the \texttt{QA} is able to compute the eigenpair corresponding to the largest eigenvalue with $100\%$ reliabilty and with improvable precision up to $\sim 10^{-8}$. As to the scalability of the algorithm,
Tab. \ref{tab:results_scala} yields a promising slightly linear increasing for growing matrix dimension, but more studies have to be performed with larger and larger $n_M$, before drawing definite conclusions.

With an eye to the future, the next challenge is to improve the  algorithm in order to  address  the generalized eigenvalue problem in its full glory, i.e.  without exploiting the non singularity of the symmetric matrix.
\begin{acknowledgements}
The Authors gratefully thank {\em Q@TN - Quantum Science and Technology in Trento}, sponsored by Universit\`a degli Studi di Trento, FBK, INFN, CNR and CINECA,
for providing  access  and run-time to the CINECA Quantum Computing facility (during the period 2021-2023), where all the numerical results shown in the present work were 
obtained by using a D-Wave  Systems quantum annealer.
A.G. acknowledges support from the DOE Topical Collaboration “Nuclear Theory for New Physics,” award No. DE-SC0023663 and Jefferson Lab that is supported by the U.S. Department of Energy, Office of Nuclear Science, under Contracts No. DE-AC05-06OR23177. During part of the development of this work A.G. was a postdoc at ECT$^*$.
T. F. thanks
the financial support from  CNPq (Conselho Nacional de Desenvolvimento Cient\'ifico e Tecnol\'ogico) grant  306834/2022-7,  CAPES (Coordena\c{c}\~ao de Aperfei\c{c}oamento de Pessoal de N\'ivel Superior), Finance Code 001, FAPESP (grant 
  2019/07767-1) and  Instituto Nacional de  Ci\^{e}ncia e Tecnologia - F\'{\i}sica
Nuclear e Aplica\c{c}\~{o}es  Proc. No. 464898/2014-5.

\end{acknowledgements}
\appendix
\section{The homogeneous Bethe-Salpeter equation}
\label{app_hBSE}
In this Appendix, some details about the physical case, which suggested the {\em non symmetric} GEVP to be investigated with the \texttt{QA},  are  briefly illustrated.  

The simplest hBSE is the one that allows to dynamically describe a  bound system composed by two massive scalars interacting through the exchange of a massive scalar (the case with a simple ladder-exchange  of a  massless  scalar between massless scalars is known as the Wick-Cutkowsky model \cite{Wick54,Cutkosky54}). In 4D Minkowski space, it reads
(see Ref. \cite{FSV2} for more details)
\bwt
\be 
\Phi_b(k,p)= G_0^{(1)}(k,p)~G_0^{(2)}(k,p) ~
\int \frac{d^4k^\prime}{(2\pi)^4}i~{\cal K}(k,k^\prime,p)\Phi_b(k^\prime,p)~,
\label{eq:BSE}\ee
\ewt
where  $\Phi_b(k,p)$ is the BS  amplitude of a two-body bound system, $p=p_1+p_2$  the total momentum of the  system, 
with  square mass $M^2=p^2$, $k=(p_1-p_2)/2$  the relative momentum, and  
$i~{\cal K}$  the interaction kernel, that
contains all the irreducible 
 diagrams \cite{BS51}. In Eq.\eqref{eq:BSE} \delete{, and} $G_0^{(i)}(k,p)$ is  the  free scalar propagator,  given by
\be
G_0^{(i)}(k,p)
=\frac{i}{(\frac{p}{2}\pm k)^2-m^2+i\epsilon}~~,
 \label{eq:G0}
\ee
with $m$ the  mass of the scalar constituents.

To proceed for obtaining actual numerical solutions is helpful to adopt the NIR \cite{nak63,Nakanishi:1971}  of the BS amplitude,  so that 
its analytic structure  is  made explicit. Within the NIR framework, the BS amplitude is written as  a proper folding of i) a non singular weight function that depends upon real variables and appears in the numerator (for the system under consideration,  one variable is compact and the other is non compact), and ii) a denominator  that contains the analytic structure. One writes  
\bwt
 \be
\Phi_b(k,p)=
~i ~\int_{-1}^1dz'
 \int_0^{\infty}d\gamma'
~
{g_b(\gamma',z';\kappa^2)\over\left[\gamma'+\kappa^2-{k}^2-p\cdot k
z'-i\epsilon\right]^{2+n}}
\label{eq:NIR}\ee
\ewt
where  $g_b(\gamma',z';\kappa^2)$ is the Nakanishi weight function (NWF), $n\ge 1$ is a suitable power that can be chosen with some degree of arbitrariness~\cite{Nakanishi:1971} (one should properly redefine the NWF) 
and 
$\kappa^2$ is given by 
 \be
\kappa^2 = m^2- {M^2\over 4}~,
\label{eq:kappa2}\ee
that is a measure of the binding, defined by $B=2m-M\ge 0$.
 Inserting such a relation  in Eq. \eqref{eq:kappa2}, one gets 
$\kappa^2 > 0$  for bound states. Interestingly, the dependence upon $z'$ of $g_b(\gamma',z';\kappa^2)$ is even
 as expected by the symmetry property of the BS amplitude for the
two-scalar system and $g_b(\gamma',z'=\pm~1;\kappa^2)=0$, as illustrated in Ref.~\cite{FSV2}. It is crucial to emphasize that once the NWF is known, then the BS amplitude can be reconstructed via Eq. \eqref{eq:NIR} and the evaluation physical observables can be carried out.

If the 4D kernel \delete{,} $i{\cal K}$ is explicitly known, one can perform the analytic integration of both sides of Eq.~\eqref{eq:BSE}. To this end, instead of using the  Minkowskian four-momentum  $k$ in Cartesian coordinates it is very useful to introduce their light-front (LF) combination, that amount to  $k\equiv\{k^\pm=k^0\pm k^3,{\bf k}_\perp\}$ (for interested readers, Ref. \cite{Dirac:1949cp} yields details on the Dirac proposal of the light-front dynamics, and in general on the relativistic Hamiltonian description of a dynamical system). 
Hence,
one gets an integral equation for the NWF, avoiding the difficulties related to the  indefinite metric of  the Minkowski space. In particular, the integral equation reads
\bwt
\be
\int_0^{\infty}d\gamma'~\frac{g_b(\gamma',z;\kappa^2)}{[\gamma'+\gamma
+z^2 m^2+(1-z^2)\kappa^2-i\epsilon]^2} =
\int_0^{\infty}d\gamma'\int_{-1}^{1}dz'\;V^{LF}_b(\gamma,z;\gamma',z')
g_b(\gamma',z';\kappa^2)\ .
\label{EQ:NIRint}\ee
\ewt
where the new kernel $V^{LF}_b$, so-called Nakanishi kernel,  is
related to the 4D BS kernel, $i{\cal K}$, in Eq. \eqref{eq:BSE}, as follows
\be
V^{LF}_b(\gamma,z;\gamma',z')=~i
p^+~\int_{-\infty}^{\infty}{d k^- \over 2\pi}~G_0^{(1)}(k,p)~G_0^{(2)}(k,p)
\nonu \times\int \frac{d^4k'}{(2\pi)^4}\frac{i{\cal K}(k,k',p)}
{\left[{k'}^2+p\cdot k' z'-\gamma'-\kappa^2+i\epsilon\right]^3}~,
\label{EQ:vbou}\ee
with $p^+=M$, once  the intrinsic frame where ${\bf p}_\perp=0$ is chosen (one adopts the rest frame for carrying out  the calculation, given our interest on the intrinsic properties of the system).

Two comments are in order. Given the  dependence on $k^2$  in both  scalar propagators and interaction kernel, one expects double poles in the Cartesian variable $k^0$. If one uses LF variables, one remains with  single poles in $k^-$ and $k^+$, obtaining a simplification in the formal treatment.
 To  numerically evaluate  the integral equation \eqref{EQ:NIRint}, one has to specify the structure of the interaction kernel. A quantitative analysis was carried out by using the exchange of one massive scalar particle,  that generates  the ladder approximation of the hBSE, by iteration. This means that the integral equation automatically produces   an infinite  number of  scalar exchanges, so that the bound-system pole of the four-leg Green's function can be  established.  The ladder kernel is given by 
\be
i{\cal K}^{(Ld)} (k,k')={i (-ig)^2 \over (k-k')^2 -\mu^2 +i\epsilon}~,
\label{eq:kernel}\ee
where $g$ is the coupling constant and $\mu$  the mass of the exchanged scalar.
A quantitative analysis of Eq. \eqref{EQ:NIRint} with the kernel in Eq. \eqref{eq:kernel}  was carried out in Ref. \cite{FSV2} by using a complete orthonormal basis for expanding the NWF $g_b(\gamma,z;\kappa^2)$. In particular, one writes
\be g^{(Ld)}_b(\gamma,z;\kappa^2) = \sum_{\ell=0}^{N_z} \sum_{j=0}^{N_{\gamma}}~
{\cal C}_{\ell j}(\kappa^2) ~G_\ell(z) ~{\cal L}_j(\gamma)\ ,
\label{eq:basis}\ee
where i) the functions $G_\ell(z)$ are given in terms of even Gegenbauer
 polynomials, $ C^{(5/2)}_{2\ell}(z)$,  by 
\be
G_\ell(z)= 4~(1-z^2) ~\Gamma(5/2)~\sqrt{{(2\ell+5/2) ~(2\ell)! \over \pi
 \Gamma(2\ell+5)}}~ C^{(5/2)}_{2\ell}(z)\ ,
 \nonu
\label{eq:bas2}\ee
and ii)
the  functions ${\cal L}_j(\gamma)$ are expressed in terms of the Laguerre polynomials,  $
L_{j}(a\gamma)$, by
\be
{\cal L}_j(\gamma)= \sqrt{a}~ L_{j}(a\gamma) ~
e^{-a\gamma/2}~~.\label{eq:bas3}\ee
The following orthonormality conditions are fulfilled
\bwt \be
\int^1_{-1}dz~G_\ell(z)~G_n(z)=\delta_{\ell n} ~~,
\quad
  \int_0^\infty~ d\gamma~{\cal L}_j(\gamma)~{\cal L}_\ell(\gamma) =
a \int_0^\infty~ d\gamma~ e^{-a\gamma}~L_{j}(a\gamma) ~L_{\ell}(a\gamma)=
~\delta_{j\ell}~.
\ee
By inserting the expansion of the NWF and approximating the kernel with only the one-scalar exchange (i.e. the ladder approximation), Eq. \eqref{EQ:NIRint} becomes
\be 
\sum_{\ell' j'} \int_0^{\infty}d\gamma \int_{-1}^{1}dz
\int_0^{\infty}d\gamma'\int_{-1}^{1}dz'
~G_{\ell}(z) {\cal L}_j(\gamma) ~{\cal B}(\gamma,z;\gamma',z';\kappa^2)~G_{\ell'}(z') {\cal L}_{j'}(\gamma') 
~{\cal C}_{\ell' j'}(\kappa^2)
\nonu
=\alpha~
\sum_{\ell' j'}
\int_0^{\infty}d\gamma \int_{-1}^{1}dz\int_0^{\infty}d\gamma'\int_{-1}^{1}dz'~G_{\ell}(z) {\cal L}_j(\gamma)
~{\cal V}^{Ld}_b(\gamma,z;\gamma',z')~
G_{\ell'}(z') {\cal L}_{j'}(\gamma')~{\cal C}_{\ell' j'}(\kappa^2).
\label{EQ:NIRint1}
\ee
where $\alpha=g^2/(32 \pi^2 )$, ${\cal B}(\gamma,z;\gamma',z';\kappa^2)$ is a function  symmetric under the exchange $\gamma \to \gamma'$ and $z\to z'$ given by
\be
{\cal B}(\gamma,z;\gamma',z';\kappa^2)={\delta(z-z')\over [\gamma'+\gamma
+z^2 m^2+(1-z^2)\kappa^2-i\epsilon]^2}~,
\ee
and ${\cal V}^{Ld}_b(\gamma,z;\gamma',z')$
 is a function non symmetric in the above exchange and is  obtained from the following integral (see Ref. \cite{FSV2})
\be 
{\cal V}^{Ld}_b(\gamma,z;\gamma',z')=~32\pi~
p^+~\int_{-\infty}^{\infty}{d k^- \over 2\pi}~
 ~
\int \frac{d^4k'}{(2\pi)^4}\frac{G_0^{(1)}(k,p)~G_0^{(2)}(k,p)}
{\left[(k-k')^2 -\mu^2 +i\epsilon\right]~\left[{k'}^2+p\cdot k' z'-\gamma'-\kappa^2+i\epsilon\right]^3}~.
\ee
\ewt
By applying the above steps, one is able 
to  formally transform the 4D hBSE into a
GEVP  with the following  structure
\be
{ A}~ {\bf v}_i= \lambda_i ~{ B} ~{\bf v}_i~, \quad {\rm with~}
i=1,2\dots~,n~,
\label{eq:gevp_app}\ee
where ${A}$ and ${ B}$ are real square matrices, calculated from rhs and lhs of Eq. \eqref{EQ:NIRint1} respectively. Moreover, the matrix $A$  is  non symmetric, while 
$B$ is symmetric,  and they have   dimension  $n=N_z\times N_{\gamma}$.  
  The eigenvector ${\bf v}_i$ ,  corresponding to the i-th eigenvalue $\lambda_i=1/\alpha_i$, contains the coefficients ${\cal C}_{\ell j}(\kappa^2)$ of the expansion of the NWF (cf. Eq. \eqref{eq:basis}). It should  be pointed out that the eigenvalues and eigenvectors can be real and complex conjugated, but we are interested only in real eigenvalues, since we aim to to describe  a physical system (recall that the eigenvalue is the inverse of the square coupling constant, in the ladder approximation we  adopted). 
\section{The binary basis}
\label{app_bb}
This Appendix illustrates how to transform  a QUBO problem in a form suitable for the evaluation with a quantum annealer. The needed step is given by the transformation  of each component of the involved {\it real }vectors into their expression in terms of a binary string,   by using $b$ {\it binary variables}. For instance,
a generic real number, $a$, such that  $|a|\le 1$, can be expressed in terms of a
binary basis with $b$ bits as follows (see also Ref. \cite{Krak:2021,krakoff2022controlled} for details)
\be  a= -q_b+ q_1~ {1\over 2}+ q_2 ~{1\over 2^2}+ q_3 ~{1\over 2^3}+~\dots~+
q_{b-1}~{1\over 2^{b-1}} ~,
\nonu
\ee
where  $a\in [-1,1)$ (see below), $q_i=0,1$ and the $b$-th bit carries the sign, i.e.   for 
$a\ge 0$ ( $a<0$) one uses  $q_b=0$ ($q_b=1$). 
In a compact form, one can rewrite $a$ as a scalar product between two vectors with $b$ 
 components, viz.
 \be a={\bf p}^T \cdot {\bf q}_b~,
 \label{eq:bb_ap} \ee
 where ${\bf q}_b\equiv \{q_1,q_2,\dots,q_b\}$ is a binary string (a column vector in our notation)   and  ${\bf p}^T \equiv\{-1,1/2,1/4,~\dots~, 1/2^{b-1}\} $ is a row vector, representing the transpose of the so-called {\em precision vector},  since $1/2^{b-1}$ controls the precision to which one wants to approximate the number $a$. 
Notice that $1$ can be reached only asymptotically, since
\be \lim_{b\to \infty}\sum_{i=1}^{b-1} {1 \over 2^i}={1\over 2} ~ \lim_{b\to \infty}
\left[{1-2^{-(b-1)}\over 1/2}\right]\to 1 ~,
\ee
while $-1$ corresponds to the binary vector $1,0,\dots~\dots,0$, truncated at any order.  Therefore, one has to  consider the right-open interval $a\in [-1,1)$.
 
One quickly generalizes the  compact expression  in Eq.~\eqref{eq:bb_ap}  to a vector ${\bf
v}$ with real components and dimension $n$, since one has
\be 
\bpmat v_1\\ v_2\\ \vdots \\v_n\epmat=
\bpmat ~{\bf p}^T  &     0    &\dots   & 0
         \\ 0      &~{\bf p}^T& \dots  & 0
         \\ \vdots &\vdots    & \ddots & 0
         \\ 0 &0    & \dots  & {\bf p}^T\epmat~\bpmat
{\bf x}_{1b}\\{\bf x}_{2b}\\ \vdots \\{\bf x}_{nb}\epmat
\nonu 
=P^T~ {\bf X}~,
\ee
where ${\bf x}_{ib}$ is the $i$-th binary vector with dimension $b$. The vector ${\bf X}\equiv\{{\bf x}_i\}$ in the last line has  dimension:  $n\times b$, and the rectangular 
matrix ${\bf P}^T$  has  $n$ rows and 
 $ n\times b$ columns.  In the last line ${\bf X}\equiv \{{\bf x}_i\}$.

Hence, for a $n\times n$ matrix $C$, and a given vector ${\bf v}$ with real components,  one can write
\bwt 
\be{\bf v}^T~C {\bf v}=\bpmat v_1& v_2& \dots &v_n\epmat~C~\bpmat v_1\\ v_2\\
\vdots \\v_n\epmat 
=\bpmat
{\bf x}^T_{1b}&{\bf x}^T_{2b}& \dots &{\bf x}^T_{nb}\epmat
\bpmat ~{\bf p} &     0    &\dots   & 0
         \\ 0      &~{\bf p}& \dots  & 0
         \\ \vdots &\vdots    & \ddots & 0
         \\ 0 &0    & \dots  & {\bf p}\epmat~C~\bpmat ~{\bf p}^T  &     0    &\dots   & 0
         \\ 0      &~{\bf p}^T& \dots  & 0
         \\ \vdots &\vdots    & \ddots & 0
         \\ 0 &0    & \dots  & {\bf p}^T\epmat~\bpmat
{\bf x}_{1b}\\{\bf x}_{2b}\\ \vdots \\{\bf x}_{nb}\epmat 
\nonu
={\bf X}^T ~PCP^T~ {\bf X}~,\ee
recalling  that ${\bf p}$ is a column vector.
\ewt
\section{Inequalities for non-symmetric-matrix eigenvalues }
\label{app_nseig}
In this Appendix, some inequalities between the eigenvalues of a $n\times n$ non symmetric matrix  and the ones of its symmetric part are  briefly reviewed, following the results of Ref. \cite{fan:1950two}. Indeed, in Ref. \cite{fan:1950two}, the most  general case of a linear transformation in a $n$-dimensional unitary space was considered, while we limit ourselves to our specific needs. 

Let us introduce the symmetric combination of a real, non symmetric matrix $A$ and its transpose $A^T$
\be
S={A+A^T\over 2} ~.
\ee
Recall that $Tr\{S\}=\sum_i \{\lambda_i(S)\} =Tr\{A\}=\sum_i \Re e\{\lambda_i(A)\} $ for real $A$, where $\lambda_i(S)$ (real) and $\lambda_i(A)$ (both real and, possibly, complex conjugated) are the eigenvalues of $S$ and $A$, with $i=1,~\dots~,n$, respectively.
 
From  Theorem II of Ref. \cite{fan:1950two}, one can deduce that the maximal eigenvalue of $S$ is an upper-bound  of the maximal eigenvalue of $A$, as sketched in what follows.
It turns out that
\be
\sum_{i=1}^q \lambda_i(S) \ge \sum_{i=1}^q\Re e\{\lambda_i(A)\}~,  ~~~~{\rm with}~~ q\le n~~,
\label{eq:Kyineq}\ee
where the two sets of eigenvalues are ordered in a decreasing way, i.e. $\lambda_1(S)\ge
 \lambda_2(S)\ge \dots~\dots\ge  \lambda_n(S)$ and the same for the
real parts of $\lambda_i(A)$. If $q=1$, then one has the following inequality
\be
 \lambda_{max}(S)\ge \Re e\{\lambda_{max}(A)\}
\label{eq:Kyineq2}\ee
The minimal eigenvalue of $S$ is  lower-bound  of the minimal eigenvalue of $A$. Recalling that $Tr\{A\}= Tr\{S\}$
one gets
\be
\sum_{i=1}^{n-1}\Re e\{\lambda_i(A)\} + \Re e\{\lambda_{min}(A)\}=\sum_{i=1}^{n-1}\lambda_i(S)+ 
\lambda_{\min}(S)~,
\nonu \ee
and  using Eq. \eqref{eq:Kyineq} with $q=n-1$ one eventually has
\be
\Re e\{\lambda_{min}(A)\}=\Biggl[\sum_{i=1}^{n-1} \lambda_i(S) -\sum_{i=1}^{n-1}\Re
e\{\lambda_i(A)\}\Biggr]
\nonu+
 \lambda_{\min}(S)\ge  \lambda_{\min}(S)~.
\ee

It should be pointed out that  the antisymmetric combination ${\cal A}=(A-A^T)/2$ leads to bounds on the
imaginary part of the eigenvalues $\lambda(A)$.

 For the sake of completeness, one can mention that if i) the largest eigenvalue of the  symmetric matrix $S$ is positive and  ii) $\lambda_{max}(A)$ is real and positive, then   $\lambda_{max}=\lambda_{max}(S)$, after applying  Eq. \eqref{eq:Kyineq2}.  In fact, one has
\begin{equation}
\|S\|_p\leq \|A\|_p=\|A^T\|_p\;,
\end{equation}
for any Schatten $p-$norm, that   becomes  a constraint between the spectral norms for $p=\infty$, i.e.
\begin{equation}
\begin{split}
\|S\|_\infty &= \max\left[|\lambda_{min}(S)|,|\lambda_{max}(S)|\right]\\
&\leq \max\left[|\lambda_{min}(A)|,|\lambda_{max}(A)|\right]=\|A\|_\infty
\end{split}
\end{equation}
From the assumed assumptions, one can writes
\be
\|S\|_\infty = \lambda_{max}(S)\leq |\lambda_{max}(A)|=\|A\|_\infty
\label{app_schat}\ee
 and after  taking into account    Eqs. \eqref{eq:Kyineq2}  one writes
\begin{equation}
\begin{split}
\Re e&\{\lambda_{max}(A)\}^2\leq \lambda_{max}(S)^2\leq |\lambda_{max}(A)|^2\\
&\quad\quad= \Re e\{\lambda_{max}(A)\}^2+\Im m\{\lambda_{max}(A)\}^2\\
\end{split}
\end{equation}
This means that if the largest eigenvalue of $A$ is real then the largest eigenvalue of $S$ will be equal to it. Unfortunately the assumptions are quite restrictive and are not fulfilled by our actual matrices.

For the symmetric combination $AA^T$, one has an analogous result.
For the sake  of simplicity (and since we are interested in), we focus on real eigenvalues   of the non symmetric matrix $A$.  One writes
 \be
A~{\bf x}_{\lambda(A)}=\lambda(A) ~{\bf x}_\lambda ~~,\quad  {\bf x}^T_{\lambda(A)} A^T= 
\lambda(A)~{\bf x}^T_{\lambda(A)}
\ee
and 
\be
{\bf x}^T_{\lambda(A)} ~A^T A~{\bf x}_{\lambda(A)}=\lambda^2(A) ~{\bf x}^T_{\lambda(A)}{\bf x}_{\lambda(A)}~.
\ee
Let us now consider the eigenvalue problem for the symmetric matrix $P=A^TA$. One has real eigenvalues and   eigenvectors generating an orthonormal basis. To establish formalism, one writes 
\be
P~{\bf x}_{\lambda_i(P)}=\lambda_i(P) {\bf x}_{\lambda_i(P)}~.
\ee
The expansion of an eigenvector of $A$ on the orthonormal basis ${\bf x}_{\lambda_i(P)}$ yields
$${\bf x}_{\lambda(A)}=\sum_i c_i~{\bf x}_{\lambda_i(P)}$$
that leads to
\be
{\bf x}^T_{\lambda(A)} ~A^T A~{\bf x}_{\lambda(A)}=\Bigl[\lambda(A)\Bigr]^2
\nonu=\sum_i \lambda_i(P) c^2_i 
|{\bf x}^T_{\lambda(A)}  {\bf x}_{\lambda_i(P)}|^2
\nonu\ge \lambda_{min}(P)~{\bf x}^T_{\lambda(A)}{\bf x}_{\lambda(A)}~,
\label{ata}\ee
 and then
 \be
\lambda_{max}(P) \ge \Bigl[\lambda(A)\Bigr]^2 \ge \lambda_{min}(P)~,
 \ee
 where an obvious inequality involving $\lambda_{max}(P)$ has been used for the lhs.
In conclusion, again the minimal(maximal) eigenvalue of the naively symmetrized matrix  is a lower(upper) bound, and a variational method can return $\lambda_{min}(P)\Bigl(\lambda_{max}(P)\Bigr)$ and not $\lambda(A)$.

Another possible approach  to be mentioned is based on the following symmetric form
 \begin{equation}
\widetilde{A}=\begin{pmatrix}
0&A\\
A^T&0
\end{pmatrix} ~.
\end{equation}
Since the non symmetric matrices $A$ and its transpose has the same set of eigenvalues but not the right, or left, eigenvectors one should solve the following {\em symmetric} GEPV
\be
\begin{pmatrix}
0&A\\
A^T&0
\end{pmatrix} ~\begin{pmatrix}
{\bf w}_i\\
{\bf v}_i
\end{pmatrix}=\lambda_i~\begin{pmatrix}
0&1\\
1&0
\end{pmatrix}~~\begin{pmatrix}
{\bf w}_i\\
{\bf v}_i
\end{pmatrix}
\ee
where $A{\bf v}_i=\lambda_i~{\bf v}_i$ and $A^T{\bf w}_i=\lambda_i~{\bf w}_i$.
Clearly, this approach doubles the dimension of the initial problem and therefore is confronted with the limits of the matrix dimension tractable by using  a quantum annealer. 

\bibliography{qctn}
\end{document}